\documentclass[twocolumn]{aastex6}

\usepackage{graphicx}
\usepackage{latexsym}
\usepackage{amsfonts,amsmath,amssymb}
\usepackage{epstopdf}
\usepackage{natbib}

\def\apjref#1;#2;#3;#4 {\par\pni\ #1,  #2, {\bf #3}, #4. \par}

\newcommand{\beq}	{\begin{equation}}
\newcommand{\eeq}	{\end{equation}}
\newcommand{\beqa}{\begin{eqnarray}}
\newcommand{\eeqa}{\end{eqnarray}}
\newcommand{\avg}[1]  {{\langle #1 \rangle}} 

\newcommand{\e}	{$^{-1}$}

\def\simlt{\lower.5ex\hbox{$\; \buildrel < \over \sim \;$}}
\def\simgt{\lower.5ex\hbox{$\; \buildrel > \over \sim \;$}}

\def\vecnabla{
              \setbox1=\hbox{$\bigtriangledown$}
                           \raise.45ex\hbox{$\bigtriangledown$\hskip-.97\wd1
                           $\bigtriangledown$\hskip-.97\wd1
                           $\bigtriangledown$\hskip-.97\wd1}
                           \raise.47ex\hbox{$\bigtriangledown$}}

\def\rsun{\ifmmode {\rm R}_{\mathord\odot}\else $R_{\mathord\odot}$\fi}
\def\msun{\ifmmode {\rm M}_{\mathord\odot}\else $M_{\mathord\odot}$\fi}
\def\lsun{\ifmmode {\rm L}_{\mathord\odot}\else $L_{\mathord\odot}$\fi}

\newcommand{\mf}		{{m_f}}

\newcommand{\mfl}		{{m_{f,\ell}}}

\newcommand{\mup}		{{m_u}}

\newcommand{\ppt}		{\psi_{p2}}

\newcommand{\scl}		{\Sigma_{\rm cl}}

\newcommand{\lacc}		{L_{\rm acc}}

\newcommand{\Ppl}		{\psi_p(L)}

\shortauthors{Gaches \& Offner}

\begin{document}
\title{A Model for Protostellar Cluster Luminosities and the Impact on the CO-H$_2$ Conversion Factor}

\author{Brandt A.L. Gaches}
\affil{Department of Astronomy, University of Massachusetts - Amherst}
\email{bgaches@astro.umass.edu}

\author{Stella S.R. Offner}
\affil{Department of Astronomy, University of Massachusetts - Amherst}
\affil{Department of Astronomy, University of Texas at Austin}
\email{soffner@astro.umass.edu}
\begin{abstract}
We construct a semi-analytic model to study the effect of far-ultraviolet (FUV) radiation on gas chemistry from embedded protostars. We use the Protostellar Luminosity Function (PLF) formalism of \cite{2011ApJ...736...53O} to calculate the total, FUV, and ionizing cluster luminosity for various protostellar accretion histories and cluster sizes. We compare the model predictions with surveys of Gould Belt star-forming regions and find the Tapered Turbulent Core model matches best the mean luminosities and the spread in the data. We combine the cluster model with the photo-dissociation region astrochemistry code, {\sc 3d-pdr}, to compute the impact of the FUV luminosity from embedded protostars on the CO to H$_2$ conversion factor, $X_{\rm CO}$, as a function of cluster size, gas mass and star formation efficiency. We find that $X_{\rm CO}$ has a weak dependence on the FUV radiation from embedded sources for large clusters due to high cloud optical depths. In smaller and more efficient clusters the embedded FUV increases X$_{\rm CO}$ to levels consistent with the average Milky Way values. The internal physical and chemical structure of the cloud are significantly altered, and $X_{\rm CO}$ depends strongly on the protostellar cluster mass for small efficient clouds. 
\end{abstract}

\section{Introduction}

In the local universe, star formation occurs exclusively within molecular clouds \citep{2007ARA&A..45..565M}. These clouds exhibit complex structure regulated by a combination of turbulence, gravity and magnetic fields \citep{2015ARA&A..53..583H}. The relative balance between these forces determines the amount of dense gas where the star formation occurs. Studying the dynamics and structure of molecular gas is paramount to understanding the star formation process. Star formation acts as a clock within molecular clouds, when internal feedback mechanisms turn on and start to impact the evolution of their natal host cloud. During star formation, knowing the dynamics is necessitated by understanding the feedback mechanisms.

Molecular clouds are composed primarily of molecular hydrogen, H$_2$. However, H$_2$ has no permanent dipole, and thus is not visible at the cold temperatures of molecular clouds. Instead, most studies rely on the emission from carbon monoxide (CO) as a proxy for total molecular gas mass. CO has the second highest molecular abundance after H$_2$, a permanent dipole and is readily excited at the temperatures and densities of molecular clouds. In addition to CO, astronomers also use a wide array of other molecules that span a range of physical and chemical conditions, including tracers of denser gas like HCN and N$_2$H$^+$ \citep{1998ApJ...504..223G, 2011MNRAS.415.1977R, 2008ApJS..175..509R, 2014ApJ...780...85V}.

Because H$_2$ is not directly observable, molecular gas mass must be determined indirectly by assuming a fixed dust-to-gas ratio or some simple relationship between H$_2$ and another molecular species. The most common conversion is X$_{\rm CO}$, which is defined as
\beq
X_{\rm CO} = \frac{N_{\rm H_2}}{W_{\rm CO}},
\label{eq:xco}
\eeq
where $N_{\rm H_2}$ is the column density of molecular hydrogen in units of cm$^{-2}$ and $W_{\rm CO}$ is the integrated intensity in K km s$^{-1}$. The typical Milky Way value is $X_{\rm CO} = 2\times10^{20}$ K km s$^{-1}$ cm$^{-2}$ \citep{2013ARA&A..51..207B}. This value implicitly assumes CO is optically thick and that molecular clouds are in rough virial equilibrium \citep{2013ARA&A..51..207B}. The related conversion factor denoted $\alpha_{\rm CO}$ relates the total CO luminosity to the molecular gas mass $M_{\rm gas}$. 

However, X$_{\rm CO}$ is subject to a variety of uncertainties.  It varies significantly within clouds \citep[e.g.,][]{pineda08}. Distance reduces the accuracy of measured CO luminosities. Outside the MW, the measured X$_{\rm CO}$  between clouds has a large dispersion, and multiple clouds may occupy an observational beam \citep{2012MNRAS.421.3127N}. It also varies with metallicity, C/O ratio, cosmic ray ionization rate and the local Far-Ultraviolet (FUV) radiation field \citep{2015MNRAS.452.2057C, 2012MNRAS.426.2142L, 2010ApJ...716.1191W, 2015ApJ...803...37B, 2006MNRAS.371.1865B, 2011MNRAS.412.1686S, 2013MNRAS.433.1223N}.
Consequently, understanding the gas chemistry and related thermal processes is crucial to interpret observations and derive accurate conversion factors.

Numerical models provide an important means to predict how abundances and gas properties vary as a function of local environment. These models range from simple one-zone models to full chemo-hydrodynamics simulations. Simple gas models \citep[i.e.,][]{1997A&A...323..953S, 2008A&A...488..623C} allow for the use of large chemical networks (hundreds of species) and parameter studies spanning diverse physical environments. Often, in these models the gas is treated as a one-dimensional, semi-infinite slab of uniform density \citep[][]{2007A&A...467..187R}. This assumption necessarily ignores the complex 3D physical structure of molecular clouds. In contrast, chemo-hydrodynamic simulations are time-intensive and, thus, restricted to smaller networks (dozens of species), but they allow for a much more accurate treatment of cloud physical conditions \citep{1997ApJ...482..796N, 2011MNRAS.412.1686S, 2011MNRAS.412..337G, 2015MNRAS.454..238W, 2016MNRAS.tmpL..19S, 2017MNRAS.465..885S}. Both approaches treat the gas as a photodissociation region (PDR) and solve chemical networks coupled to the physical environment. 

By convention, the FUV radiation field is assumed to be a one-dimensional,  monochromatic flux incident on the cloud boundary, which represents the interstellar radiation field (ISRF). This treatment implicitly assumes that only external stellar sources influence the cloud chemistry. However, forming stars radiate their environment, producing chemical changes deep within the cloud. Protostellar radiation is produced by both accretion and stellar processes, such that embedded sources often have luminosities much higher than that of main sequence stars of the same mass \citep{2014prpl.conf..243K, 2011ApJ...740...74K, 2009ApJ...703..131O, 2014MNRAS.437...77B}. The protostellar spectrum includes radiation at FUV wavelengths and, for high-mass stars, ionizing radiation. Therefore, once molecular clouds begin forming stars the local radiation field is set by both the ISRF and radiation from embedded star formation. 

To date, no PDR studies have directly included 
embedded sources. Instead, some recent theoretical work indirectly modeled how the star formation rate (SFR) affects X$_{\rm CO}$. \cite{2010ApJ...720..226P} studied the physio-chemical nature of high-density star formation systems, such as ULIRGS. They derived a correlation between the supernova rate (and hence star formation rate) and the galactic average FUV background and cosmic ray ionization rate. They found that while the FUV radiation is quickly attenuated, cosmic rays are able to penetrate and heat the entire cloud. \cite{2015ApJ...803...37B, 2017ApJ...839...90B} used one-zone models to study the destruction of CO by cosmic rays across a parameter space spanning many different types of galaxies. \cite{2015MNRAS.452.2057C} combined the \cite{2010ApJ...720..226P} model with hydrodynamic simulations and post-processing to study the impact of the star formation rate on X$_{\rm CO}$. They found that X$_{\rm CO}$ increased with the star formation rate. However, none of these studies included embedded radiation or cosmic rays from protostars. 

In this paper, we formulate a simple cloud model that includes internal sources of FUV radiation in order to study variations in CO chemistry as a function of star formation activity. Section \ref{sec:model} describes the semi-analytic model we use to calculate the cluster luminosities and our astrochemistry method. Section \ref{sec:res} shows the results of the calculations for two different physical models: one where the cloud gas mass is fixed and a second where the cloud gas mass is varied as a function of star formation efficiency. Section \ref{sec:dis} discusses the implications of our study for observations and compares the results to prior work.

\section{Modeling the CO Emission of Star-Forming Clouds}\label{sec:model}

\subsection{Star Cluster Model}

We summarize the Protostellar Luminosity Function (PLF) formalism from \cite{2011ApJ...736...53O} here for completeness and discuss our extensions to the work.

The PLF is derived by adopting an accretion model, which in turn prescribes the underlying distribution of protostellar masses assuming that the final masses of the protostars obey a specified stellar initial mass function (IMF). In this framework, the accretion rate of a particular protostar, $\dot{m}$, is solely a function of its current mass, $m$, and its final mass, $\mf$. The Protostellar Mass Function (PMF) describes the distribution of current protostellar masses, i.e., the present-day protostellar mass function.
\cite{2010ApJ...716..167M} define the PMF as:
\beq
\psi_p(m)=\int_\mfl^\mup \ppt(m,\mf) d\ln\mf,
\label{eq:pmf}
\eeq
where $\ppt(m, \mf)$ is the bi-variate PMF which defines the fraction of protostars in a star-forming region with current masses in the range $dm$ and final masses in the range $d\mf$. The bi-variate PMF is related to the bi-variate number density, $dN^2_p$, within a cluster by:
\beq
dN^2_p = N_p \psi_{p2}(m, m_f) d \ln m d\ln m_f,
\label{eq:pmf2}
\eeq
where $N_p$ is the number of protostars in the cluster. We denote the stellar IMF as $\Psi(\mf)$. 
For a steady star formation rate,
\beq
\ppt(m,\mf)=\frac{m\Psi(\mf)}{\dot m\avg{t_f}},
\label{eq:ppt}
\eeq
where $\Psi(\mf)$ is the stellar IMF, $t_f$ is the time it takes to form a star with mass $\mf$ and $\avg{t_f}$ is the average time to form a star:
\beq
\avg{t_f} = \int\limits_{m_l}^{m_u} d\ln \mf \Psi(\mf) t_f(\mf).
\eeq
Following \cite{2010ApJ...716..167M}, we assume $\Psi(\mf)$ is a Chabrier IMF \citep{2005ASSL..327...41C} truncated at some maximum mass, $m_u$. 

\cite{2011ApJ...736...53O} parameterize the accretion model as:
\beq
\dot{m} = \dot{m}_1 \left (\frac{m}{\mf} \right )^j \mf^{j_f} \left [ 1 - \delta_{n1}\left ( \frac{m}{\mf} \right )^{1-j} \right ]^{1/2},
\label{eq:macc}
\eeq
where $\dot{m}_1$ is a constant, $j$ and $j_f$ are model parameters, and $\delta_{n1}$ is a parameter determining whether the accretion rate limits to zero at $t_f$ (``tapered"). In this study, we consider three different accretion histories:
\begin{enumerate}
\item Inside out collapse of an Isothermal Sphere (IS) \citep{1977ApJ...214..488S}, which gives
\beq
\dot m = \dot m_{\rm IS} = 1.54\times10^{-6} (T/10 \, K)^{3/2} ~~~M_\odot~\mbox{yr\e},
\eeq
where $T$ is the gas temperature. In this model, the accretion rate is constant for a given temperature and is independent of stellar mass.  
\item Turbulent Core (TC) model \citep{2003ApJ...585..850M} in which the turbulent pressure exceeds thermal pressure. The accretion rate is 
\beq
\dot m_{\rm TC}=3.6\times 10^{-5}\scl^{3/4}\left(\frac{m}{\mf}\right)^j \mf^{3/4}
~~~M_\odot~\mbox{yr\e},
\label{eq:mdtc}
\eeq
\t where $\scl$ is the surface mass density, given in units of ${\rm g \, cm^{-2}}$, and m and m$_f$ defined above.
$\dot{m}_1 = \dot{m}_{\rm TC} = 3.6\times 10^{-5}\scl^{3/4}$. 
Following \cite{2003ApJ...585..850M}, we use  $j = \frac{1}{2}$.  In this model, higher mass stars accrete at higher rates.
\item Tapered Turbulent Core (TTC) model \citep{2011ApJ...736...53O}
\beq
\dot m_{\rm TTC} = \dot m_{\rm TC} \left [ 1 - \left (\frac{m}{m_f} \right )^{1-j} \right ]^{1/2}
~~~M_\odot~\mbox{yr\e}
\eeq
\t where the parameters are taken to be the same as the turbulent core model but $\delta_{n1} = 1$. The tapered accretion rate produces smaller luminosities in later stages of protostellar evolution. 
\end{enumerate}
For accretion histories formulated in this way, the formation time of an individual star is:
\beq
t_f = t_{f1} m_f^{1 - j_f} (1 + \delta_{n1}),
\eeq
where
\beq
t_{f1} = \frac{1}{(1-j)\dot{m}_1}
\eeq
and $t_{f1}$ is the time to form a star of $1~M_\odot$. We discuss the impact of adopting a different tapering model in Appendix \ref{appendix:tappar}.

The PLF, $\Ppl$, is defined such that $\Ppl d\ln L$ is the fraction of protostars within the luminosity range $dL$.  \cite{2011ApJ...736...53O} showed that the bi-variate PLF is related to the bi-variate PMF by
\beq
\ppt(L,\mf) d\ln L\, d\ln \mf=\ppt(m,\mf) d\ln m\, d\ln\mf,
\label{eq:Ppt}
\eeq
such that the PLF is defined
\beq
\Ppl = \int d\ln m \cdot \ppt(L,m).
\label{eq:ppl1}
\eeq
\cite{2011ApJ...736...53O} calculate the PLF by transforming Equation \ref{eq:ppl1} to:
\beq
\Ppl = \int\limits_{m_{f,l}(L)}^{m_u} d\ln \mf \frac{\psi_{p2}(m(L), \mf)}{\left | \frac{\partial L}{\partial m} \right | },
\label{eq:pplt}
\eeq
where $m_{(f,l)}(L)$ = max($m_l$, $m(L)$). 
 
To calculate the luminosities,  we adopt the model in \citet{2009ApJ...703..131O}, which is based on \cite{2003ApJ...585..850M}. This model represents the protostellar luminosity as the sum of two parts, $L = L_{\rm acc} + L_{\rm int}$, where $L_{\rm acc}$ is the accretion luminosity and $L_{\rm int}$ is the internal protostellar luminosity, including Kelvin-Helmholz contraction and nuclear burning.
The total accretion luminosity is defined by:
\beq
\lacc= f_{\rm acc} \frac{G m \dot{m}}{r},
\eeq
where $f_{\rm acc}$ is the efficiency at which mechanical energy is converted  to radiation, $G$ is the gravitational constant, $m$ is the protostar mass, $\dot{m}$ is the accretion rate (given by Equation \ref{eq:macc}) and $r$ is the protostar radius calculated following \cite{2009ApJ...703..131O}. Following \cite{2011ApJ...736...53O}, we use $f_{\rm acc} = 0.75$. The total internal luminosity is approximated by the main sequence mass - luminosity relationship given in \cite{1996MNRAS.281..257T}:
\beq
L_{\rm int} = \frac{\alpha M^{5.5} + \beta M^{11}}{\gamma + M^3 + \delta M^5 + \epsilon M^7 + \zeta M^8 + \eta M^{9.5}}.
\eeq

Coupling this model to our PDR calculation requires some assumption about the shape of the protostellar spectrum. We assume that each luminosity component is a blackbody as described by the Planck function, such that the luminosity in a given energy range is
\beq
L_{\rm \Delta E} = f_{\rm \Delta E}(L_{\rm acc}) \times  L_{\rm acc} + f_{\rm \Delta E}(L_{\rm int}) \times L_{\rm int},
\label{eq:lcalc}
\eeq
where $f_i(L)$ is the fraction of the Planck function within the given energy range of interest. The blackbody temperature is derived using the Stefan-Boltzmann law with the protostar radius and luminosity from the \citet{2009ApJ...703..131O} model. In the limiting case where $\Delta E \rightarrow \infty$, $L_{\rm \Delta E} \rightarrow L$.

This model is intended to represent relatively young clusters, whose membership is dominated by protostars. Appendix \ref{appendix:msstars} discusses the results for clusters which include a secondary population of main sequence stars.

\subsection{Statistical Sampling}

\begin{figure}
\plotone{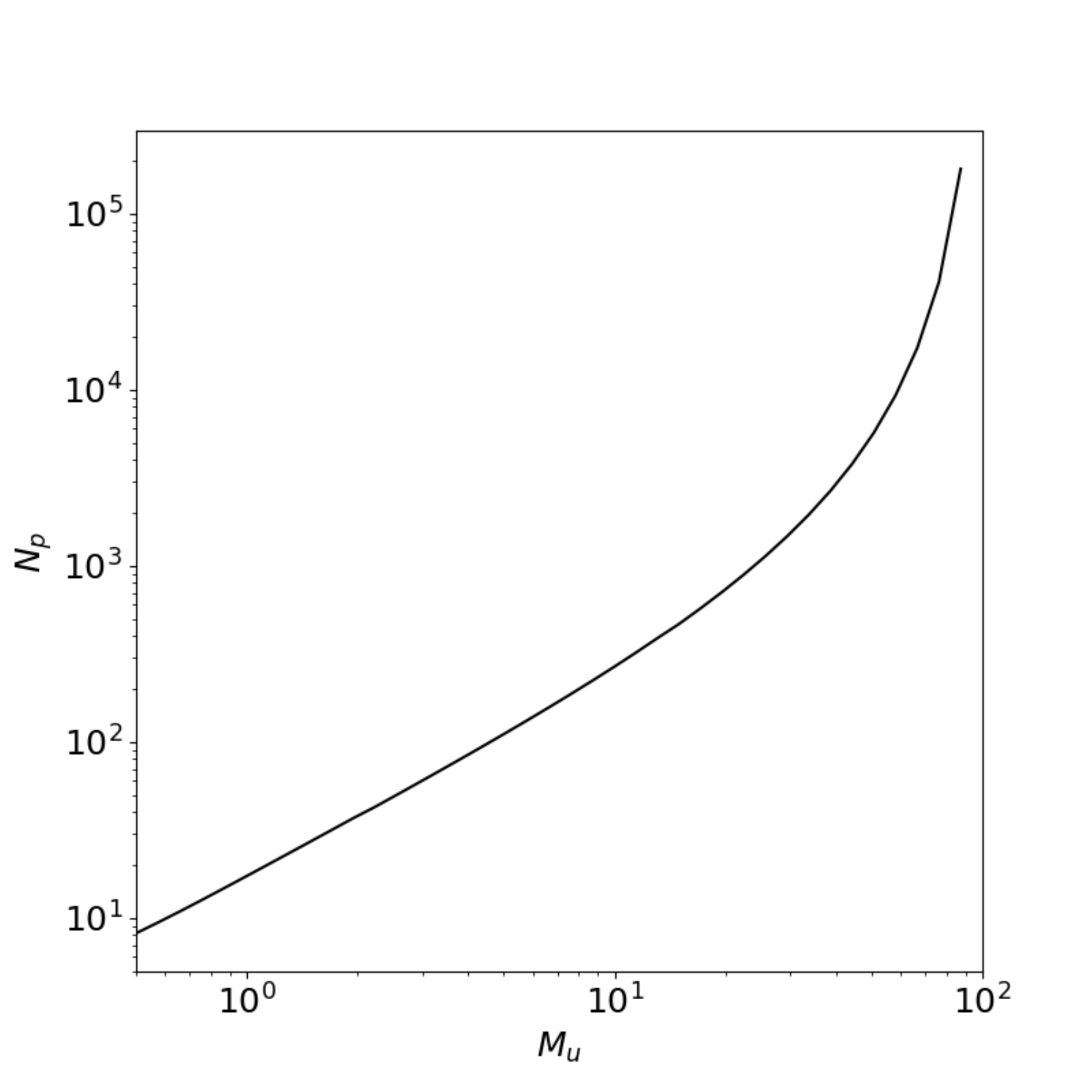}
\caption{\label{fig:nfm} Number of stars as a function of the highest mass star in the cluster. }
\end{figure}

The PMF describes the likelihood that a cluster contains a protostar with a specific instantaneous mass and final mass. Because lower mass stars are much more numerous than higher mass stars,  small clusters are statistically unlikely to include any high-mass stars. Under the assumption of a perfectly sampled PMF, the number of stars in a cluster, $N_p$, can be related to final mass of the highest mass star within the cluster, $m_u$. \cite{2010ApJ...716..167M} show that the cluster size, highest mass star in the cluster and the maximum possible stellar mass, $m_{\rm max}$ are related by:
\beq
\frac{1}{N_p(m_u)} = \int \limits_{m_u}^{m_{\rm max}} \psi_p(m) d \ln m.
\label{eq:Nmu}
\eeq
From an observational stand-point, the maximum mass, $m_{\rm max}$, is highly uncertain due to a variety of factors. Crowding in clusters and unresolved binarity make measurements of individual high-mass stars challenging \citep{2014prpl.conf..149T}. Furthermore, constraining $m_{\rm max}$ requires measuring the populations of very young massive clusters, which are rare and distant. This work focuses mainly on small to intermediate clusters ($N_p\sim 10-10^5$), so  we adopt $m_{\rm max} =100 M_{\odot}$. The total cluster mass is then M$_{\rm cl}$ = N$_p \times \avg{m}$, where $\avg{m} = \int\limits_{m_l}^{m_{\rm max}} d \ln m m \Psi(m)$. Figure \ref{fig:nfm} shows $N_p(m_u)$ as a function of the highest mass star in the cluster. We adopt a minimum mass, $m_{\rm min} = 0.033$ M$_{\odot}$. For the TTC model, the average mass $\avg{m} \approx 0.2$ M$_{\odot}$.

Equation \ref{eq:pplt} can be numerically integrated given a protostellar model for $L(m)$ and $r(m, \mf)$. This approach allows the distribution to be calculated exactly, i.e., direct integration produces perfect sampling of the underlying function.  However, the stellar radius undergoes several discontinuous jumps due to changes in the nuclear state \citep[e.g., Fig.~5 in][]{Offner14} and is consequently difficult to invert. Moreover, the mass functions of small clusters are subject to Poisson statistics and, thus, not perfectly sampled. We therefore adopt a statistical approach to compute the PLF and cluster properties.

\begin{figure}
\plotone{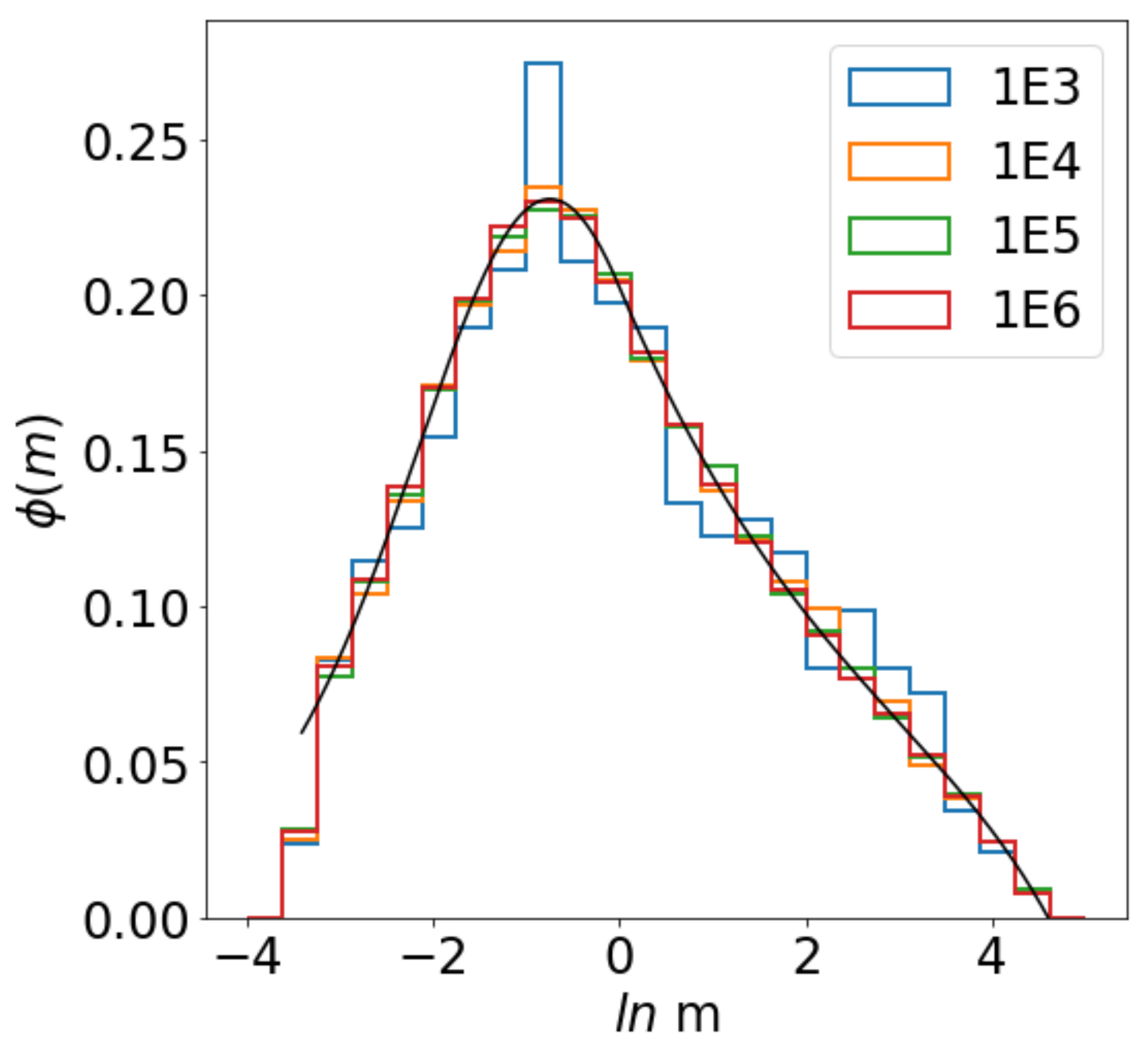}
\caption{\label{fig:conv} Protostellar Mass Function as a function of the logarithm of the protostellar mass for the isothermal sphere accretion model. The different colored histograms represent different distributions from the indicated number of protostars in the legend. The black line indicates the analytic PMF calculated integrating Equation \ref{eq:pmf} directly.}
\end{figure}

We calculate the PLF and PMF of a cluster using the conditional probability method. The first step of the method is to marginalize the bivariate PMF (\ref{eq:ppt}) over the protostar final mass, $m_f$. This one-dimensional distribution function is then sampled for a protostar mass, $m$ using the inversion method numerically. We then calculate the conditional probability distribution for the final mass given the current mass, $\psi(m_f|m) = \frac{\psi_{\rm p2}(m=m, m_f)}{\psi_p(m=m)}$. The conditional probability is then sampled using the inversion method again to obtain the final mass, $m_f$. This procedure is done for as many protostars as in each cluster. The protostellar masses drawn this way converge to the analytic PMF with a sample of 10$^5$ protostars. Figure \ref{fig:conv} shows the convergence of the PMF distribution to the analytic result for the isothermal sphere accretion model as a function of the number of stars included in the distribution. We find that the distribution converges well to the analytic distribution by $N_* \approx 10^5$.

To calculate cluster statistics, we draw $N_*$ protostars for a number of mock clusters, $N_{\rm cl}$, using the procedure described above. For each mock cluster, we calculate the bolometric, FUV and ionizing luminosities for each protostar using Equation \ref{eq:lcalc}. The total luminosities and masses are calculated for the mock cluster. After drawing N$_{\rm cl}$ mock clusters, we calculate the average and spread of the different total luminosities and the mass. When we compare to observations in Section \ref{sec:res} 
to achieve statistical robustness for the mean and the spread. For the chemistry, we use the average of the total cluster luminosities and masses. As such, we optimize the procedure by calculating the running mean of the total bolometric luminosity and drawing clusters until the running mean converges to 0.1\% relative error. We find that the running mean converges in $N_{cl} \approx 15-20$ across 4 dex of N$_*$.

\subsection{PDR Chemistry}

We use the photo-dissociation region code {\sc 3d-pdr}\footnote{https://uclchem.github.io/} \citep{2012MNRAS.427.2100B} to model the chemistry of the molecular gas in our models. {\sc 3d-pdr} obtains the gas temperature and abundance distributions for a given input density distribution by balancing the heating and cooling. Cooling mainly occurs due to [CI], [OI] and [CII] forbidden line emission. {\sc 3d-pdr} includes four heating mechanisms: i) photoelectric heating of dust grains due to FUV radiation, ii) de-excitation of vibrationally excited H$_2$,  iii) cosmic-ray heating of the gas and iv) heating due to turbulent dissipation. {\sc 3d-pdr} also requires the strength of the incident radiation field, information about any embedded sources, the cosmic ionization rate and the gas velocity dispersion. See \citet{2012MNRAS.427.2100B} for further technical details.  We adopt the {\sc umist12} chemical reaction network \citep{mcelroy13}, which uses 215 species and follows approximately 3,000 reactions. We use the initial atomic abundances in Table \ref{tab:abund} from \cite{2000ApJ...528..310S}. By construction,  the gas is initially entirely atomic and neutral.

\subsection{Cloud Model}

Each molecular cloud is represented by  a one-dimensional slab of constant density. The depth of the cloud
is determined by the total molecular gas mass, 
\begin{equation} 
R_c = \left( \frac{3 M_{\rm gas}}{4 \pi n \mu m_p} \right)^{1/3},
\end{equation}
where $m_p$ is the mass of a proton, $n$ is the gas number density and $\mu$ is the mean molecular weight, taken to be $\mu = 1.4$ since the cloud is assumed to be initially atomic and neutral (see below). The total gas mass is set according to two different gas models as described below. 

In these models, there are two FUV components: an external field, $F_{\rm ext} = 1$ Draine \citep{draine78}, and an internal field $F_{\rm src}$, from embedded sources as given by the average cluster FUV luminosity from the mock clusters. We scale the latter to the Draine field by renormalizing the units by $\chi_0 =  1.7 G_0$ where $G_0 = 1.6\times10^{-3}	$ erg s$^{-1}$ cm$^{-2}$ is the Habing field \citep{1968BAN....19..421H}. We adopt the fiducial cosmic ray ionization rate from \cite{2006MNRAS.371.1865B} of $\xi_0 = 1.3\times10^{-17}$ s$^{-1}$  per H$_2$ molecule. Previous studies of H$_3^+$ chemistry \citep{2007ApJ...671.1736I} and H$_n$O$^+$ chemistry \citep{2015ApJ...800...40I} towards diffuse clouds find larger cosmic ray ionization rates on the order of 10$^{-16}$. However, there is a large spread in observed values, and the cosmic ray ionization rate appears to decrease towards clouds with higher column density \citep{2009A&A...501..619P}. We study the implications of cosmic ray ionization rates higher than the fiducial value in Section \ref{sec:res}.

\begin{figure}
\plotone{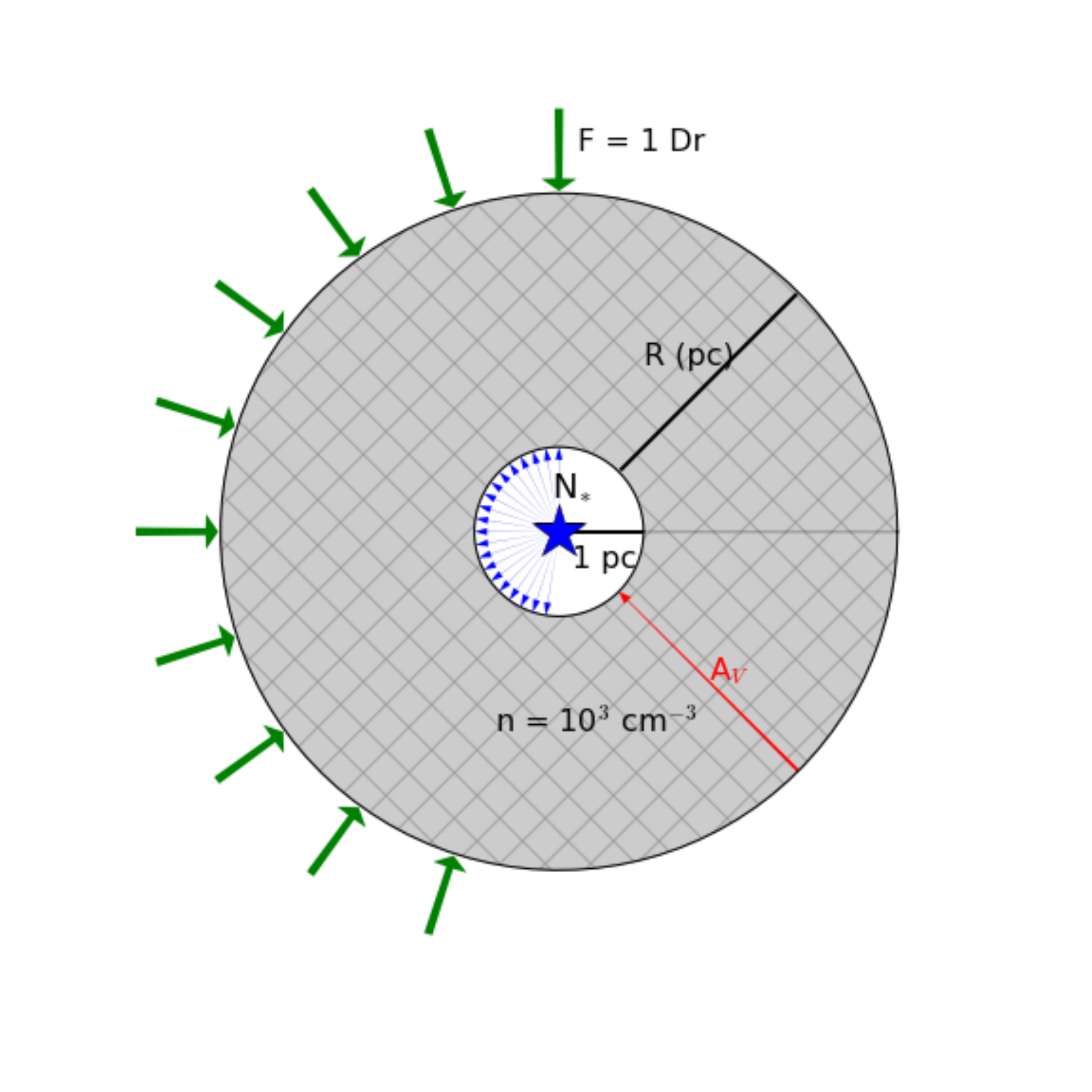}
\caption{\label{fig:schematic} Schematic of the geometry assumed in our cloud models. The number of stars in the cluster is $N_*$ and the radius of the cloud is calculated assuming the gas has a constant density. The external and internal fluxes are isotropic, with only half the arrows being shown for clarity.}
\end{figure}

Figure \ref{fig:schematic} displays a schematic of our cloud model. The field from embedded sources is indicated by blue arrows and the external field is represented by green arrows. We define $A_V$, the dust extinction through the cloud, such that the surface has $A_V = 0$ and the stars are located at high $A_V$ in the cloud center. We place the cluster within an evacuated bubble to approximate the effects of feedback mechanisms. The bubble has a size $R_{\rm bubble}$ given by
\beq
R_{\rm bubble} = {\rm max}(1\, {\rm pc}, R_s)
\eeq
where $R_s$ is the Str\"{o}mgren sphere radius for the given density and ionizing luminosity from the cluster model
\beq
R_s = \left ( \frac{3 Q_0}{4\pi \alpha_B n_e^2} \right )^{\frac{1}{3}}
\eeq
where we approximate $Q_0 = \frac{L_{\rm Ionizing}}{18 \,{\rm eV}}$ following \cite{2011piim.book.....D} for first-order computation, $\alpha_B$ is the recombination case B coefficient, and we assume $n_e \approx n_H$.

In addition to the abundances, {\sc 3d-pdr} computes the line emissivities for C, C$^+$ and CO assuming  non-local thermodynamic equilibrium (NLTE) and accounting for optical depth. We use these line emissivities to calculate the CO (1-0) integrated line intensity  following \cite{2007A&A...467..187R}:
\beq
I = \frac{1}{2\pi} \int\limits_{0}^{R} j \, dz ~~~{\text{ (erg s$^{-1}$ cm$^{-2}$)}},
\eeq
where
\beq
W = \frac{c^3}{2 k_b \nu^3}\, I ~~~{\text{(K  km s$^{-1}$)}}
\eeq
and $c$ is the speed of light, $k_b$ is the Boltzmann constant and $\nu = 115.3$ GHz is the frequency of the CO (1-0) line. We calculate the H$_2$ column density directly from the {\sc 3d-pdr} abundances
\beq
N(H_2) = \int\limits_{0}^{R} n(H_2) \,  dz ~~~({\rm cm^{-2}}),
\eeq
where $n(H_2) = n_{\rm gas}X(H_2)$ and $X(H_2)$ is the H$_2$ abundance.

\begin{deluxetable}{cc}
\tablecolumns{2}
\tablecaption{Atomic Abundances \label{tab:abund}}
\tablehead{\colhead{Species} & \colhead{Abundance Relative to H}}
\startdata
H & 1.0\\
He & 0.1\\
C & 1.41$\times10^{-4}$\\
N & 7.59$\times10^{-5}$\\
O & 3.16$\times10^{-4}$\\
S & 1.17$\times10^{-5}$\\
Si & 1.51$\times10^{-5}$\\
Mg & 1.45$\times10^{-5}$\\
Fe & 1.62$\times10^{-5}$
\enddata
\tablecomments{Atomic abundances adopted from \cite{2000ApJ...528..310S}.}
\end{deluxetable}

\subsection{A Coupled Cluster and PDR Model}
\begin{deluxetable*}{c|cccccc}
\tablecolumns{7}
\tablecaption{Chemical Models \label{tab:models}}
\tablehead{\colhead{Model Name} & \colhead{Constant Mass} & \colhead{Constant Efficiency} & \colhead{Velocity Dispersion} & \colhead{Density (cm$^{-3}$)} & \colhead{$\xi$} & \colhead{Internal Sources}}
\startdata
\label{model:1}CM\_1000D\_1kms\_1$\xi$ & $\checkmark$ &  & 1 km$/$s & 1000 & $\xi_0$ & $\checkmark$ \\
\label{model:2}CE\_1000D\_V\_1$\xi$ & & $\checkmark$  & Virial & 1000 & $\xi_0$ & $\checkmark$ \\
\label{model:3}CE\_500D\_V\_1$\xi$ & & $\checkmark$ & Virial & 500 & $\xi_0$ & $\checkmark$ \\
\label{model:4}CE\_1000D\_1kms\_1$\xi$ & & $\checkmark$ & 1 km$/$s & 1000 & $\xi_0$ & $\checkmark$ \\
\label{model:5}CE\_1000D\_V\_100$\xi$ & & $\checkmark$ & Virial & 1000 & 100$\xi_0$ & $\checkmark$ \\
\label{model:6}CE\_1000D\_V\_1$\xi$\_NS & & $\checkmark$ & Virial & 1000 & $\xi_0$ &  \\
\enddata
\tablecomments{Names and parameters for the different chemical models used. Virial denotes the velocity is calculated using Eq. \ref{eq:virial}. } 
\end{deluxetable*}

We use two different models for the total gas mass and cloud velocity dispersion. The first, denoted by CM, is a constant-mass model where the total gas mass is $M_{\rm gas} = 10^4 \, M_{\odot}$. This model also assumes a constant velocity dispersion of 1 km s$^{-1}$, making it slightly sub-virial. The second model, denoted by CE, is a constant efficiency model where the total gas mass depends on the stellar mass: $M_{\rm gas} = \frac{M_*}{\varepsilon_g}$, where $\varepsilon_g$ is related to the star formation efficiency: 
\beq
\varepsilon_{\rm tot} = \frac{M_*}{M_{\rm gas} + M_*} = \frac{\varepsilon_g}{\varepsilon_g + 1}. 
\eeq
We vary $\varepsilon_g$ between 0.01 and 0.2, or $\varepsilon_{\rm tot}$ between 0.01 and 0.166. This produces total gas masses from 10$^3$ M$_{\odot}$ to 10$^8$ M$_{\odot}$. We calculate the velocity dispersion for the constant efficiency models assuming the clouds are in virial equilibrium, such that
\beq
\sigma_v = \left (\frac{4 \pi G}{15}\right )^{1/2} R \rho^{1/2},
\label{eq:virial}
\eeq
where $G$ is the Gravitational constant. 

Table \ref{tab:models} summarizes the six models we consider. The fiducial CM model is denoted CM\_1000D\_1kms\_1$\xi$ and the fiducial CE model is denoted CE\_1000D\_V\_1$\xi$.
We include a model with a lower number density of 500 cm$^{-3}$ (500D), models that vary and fix the velocity dispersion (V and 1~kms, respectively), one model with enhanced cosmic ray ionization rates, and a model without internal sources.  This last model allows us to compare the influence of stellar sources relative to the external field.

\section{Results}\label{sec:res}

\subsection{Cluster Luminosities and Comparison with Local Milky Way Regions}

We compare our PLF cluster model to data from three recent surveys of molecular clouds. \cite{2013AJ....145...94D} and \cite{2012AJ....144...31K} each survey a number of well-studied clouds located in the Gould Belt.  \cite{2014AJ....148...11K} present a survey of the Cygnus X region,  which is 1.4 kpc away and is one of the most massive star-forming complexes within 2 kpc of the Sun. Cygnus X contains multiple evolved OB associations with dozens of O stars and hundreds of B stars. It is also the largest cluster in our comparison with nearly 2,000 identified protostars.

The surveys adopt slightly different conventions for identifying protostars. \cite{2013AJ....145...94D} define protostars as point sources with at least one detection at $\lambda \ge 350 \mu{\rm m}$. They argue this constraint removes older, non-protostellar sources, while including only sources that are still deeply embedded in dusty envelopes. \cite{2012AJ....144...31K} use color magnitude diagnostics to identify protostellar sources and do not require a sub-millimeter detection. Both surveys thus have their own biases: \cite{2013AJ....145...94D} likely underestimate the number of dim sources, since protostars embedded in very low-mass cores, which fall below the sub-millimeter detection limit, are excluded. \cite{2012AJ....144...31K} possibly over-estimates the number of low-luminosity sources, by including older, less embedded sources that would have been filtered out by requiring a sub-millimeter detection. In Chameleon II, however,  \cite{2012AJ....144...31K} excludes some of the objects found in \cite{2013AJ....145...94D}. The net effect is that clusters reported in \cite{2012AJ....144...31K} tend to have have larger populations of low-luminosity sources \cite{2013AJ....145...94D}. Additional disagreement occurs because the two surveys assume different distances for a few of the shared clouds (i.e., for Perseus the former uses a distance of 230 pc and the later uses 250 pc).  An order of magnitude luminosity discrepancy is evident between the two surveys for Chameleon II because the selection criteria in \cite{2012AJ....144...31K} only has one of the three Chameleon II objects in \cite{2013AJ....145...94D}. Both of the surveys likely suffer from incompleteness at low luminosities to some degree due to missing sources that are either very low-mass ($m_* \lesssim 0.2 M_\odot$, \citealt{2011ApJ...736...53O}) or very young and embedded ($L \le 0.1 L_\odot$, e.g., \citealt{2017arXiv171002506M})\ 

Given that a significant number of dim protostars could lie below the survey detection limits, we assume the reported number of sources in both cases is a lower limit that underestimates the true number by up to a factor of 2. This conservative completeness assumption encompasses sources that are either very low mass, e.g., $\lesssim 0.1 M_\odot$, or are undergoing a period of low accretion. \cite{2008ApJ...684.1240E} cite a 50\% completeness limit for the Bolocam 1.1 mm survey, and we use that as an upper limit in the error of the observed protostar number counts to account for incompleteness. Furthermore, while our derived PLF luminosity value is exact, the measured bolometric luminosities have some intrinsic uncertainties that are not reported. 

\begin{figure*}
\centering
\includegraphics[width=\textwidth]{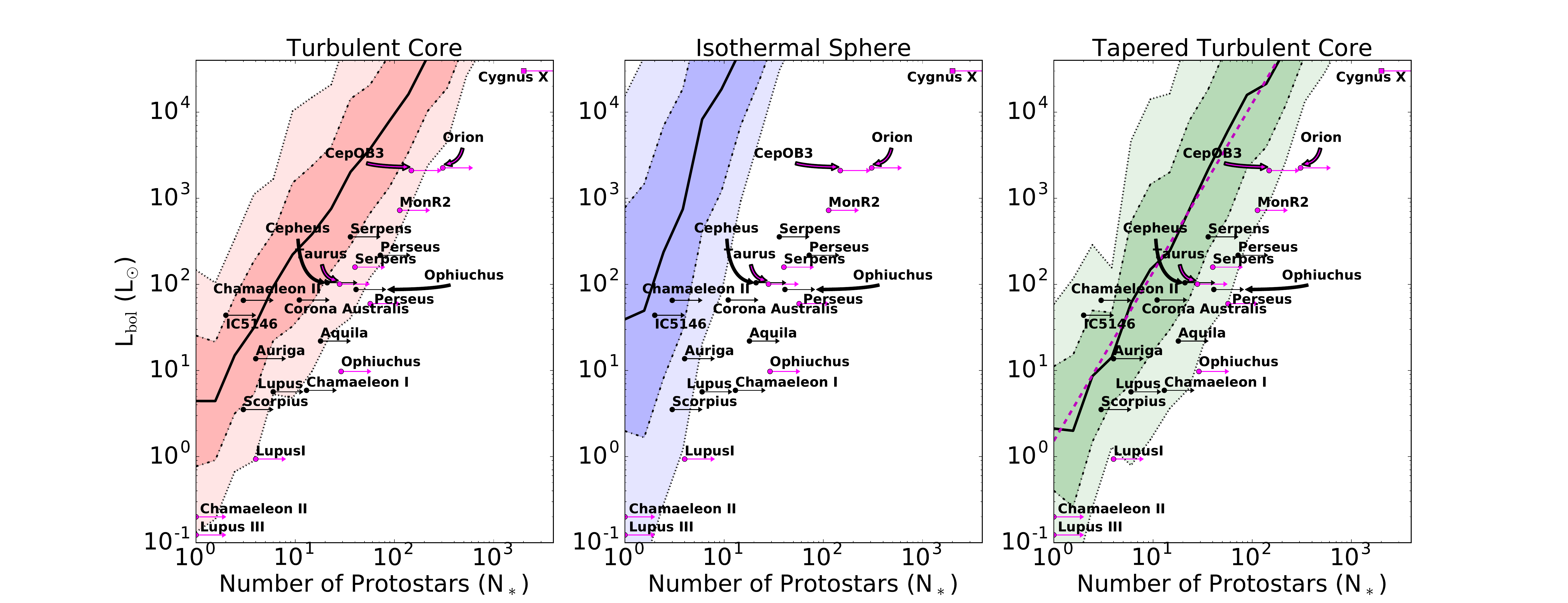}
\caption{\label{fig:clusterL} Total cluster luminosity as a function of the number of protostars for three different accretion histories. The black solid lines indicate the mean of the luminosity distributions, $\avg{L}$. The dark and light colored bands indicate the 1 and 2 $\sigma$ spread of the distribution, respectively.  The magenta dotted line is the best fit for the TTC model (Equation \ref{eq:Lbol}). The black data points indicate the sum of the bolometric luminosities for each cluster in \cite{2013AJ....145...94D}. The pink circles show clusters from the \cite{2012AJ....144...31K} catalog and the pink square is Cygnus X from \cite{2014AJ....148...11K}. The arrows indicate that each of the points are likely lower limits to the actual number due to incompleteness at low luminosities.} 
\end{figure*}

Figure \ref{fig:clusterL} shows the model predictions for the total cluster luminosity across three orders of magnitude in cluster size. We include predictions for the three different accretion models described above. The figure shows the mean total luminosity of the statistically sampled clusters, where dotted lines indicate the one and two $\sigma$ deviations from the mean. These boundaries are slightly irregular since they are influenced somewhat by the statistical sampling of the mock clusters. For smaller clusters, a broader PMF creates a correspondingly large spread in the cluster luminosity. The spread decreases for large clusters as the PMF becomes well-sampled. For all three models, the total luminosity scales superlinearly with cluster size until N$_* \approx 10^3$ when it approaches a linear scaling. For the TTC model, the bolometric luminosity is fit by:
\begin{equation}\label{eq:Lbol}
\log L_{\rm Bol} = 
\begin{cases}
1.96 \cdot\log N_* + 0.18 & \text{$\log N_* < 2.78$} \\
\log N_* + 5.63 & \text{$\log N_* \ge 2.78$}
\end{cases}
\end{equation}

Inspection of Figure \ref{fig:clusterL} shows that the IS model agrees poorly with the data. It fails to match both the mean and the spread of observed luminosities of the low-mass clusters. However, both the tapered and non-tapered TC models are able to reproduce the observed spread quite well. This suggests that poor statistical sampling together with a significant range of underlying accretion rates is needed to explain the observational data.  All the models appear to significantly over-predict the total luminosity of Cygnus X. However, the brightest sources are saturated in the MIPS 24 $\mu$m band, and their luminosities are under-estimated in the catalog (R.~Gutermuth priv.~comm.).

The TC model does a good job of representing the spread as a function of cluster size, but it over-predicts the luminosities of clusters with sizes $N_* = 10-100$, where observed data points fall outside the $2\sigma$ statistical sampling error. The TTC model does exceptionally well in encapsulating the data from all the surveys. The majority of the observed cluster bolometric luminosities are included within the 2$\sigma$ spread of the model predictions. All models over-predict the luminosities at the smallest cluster sizes. The discrepancy may be caused by several factors, such as completeness limits and differences in the physical parameters we assume, which we discuss in more detail in Appendix \ref{appendix:facc}. As a result of this comparison, we adopt TTC as the fiducial accretion model for the analysis in the following sections.

While the total bolometric luminosity is an observable quantity and, thus, useful for evaluating the accuracy of PLF predictions, our PDR calculations require the strength of the FUV radiation field as an input. Since protostellar radiation is heavily reprocessed by the surrounding dusty envelope, it is not possible to directly measure the FUV component of the spectrum. Instead, our PLF models provide an approach to calculate the fraction of short-wavelength radiation. We use the approximation in Equation\ref{eq:lcalc} to calculate the FUV and ionizing luminosity for each protostar in a given cluster and then compute the total by summing over all protostars, i.e., $L_{\Delta E} = \sum\limits_i L_{\Delta E}^i$. 

\begin{figure*}
\centering
\includegraphics[width=\textwidth]{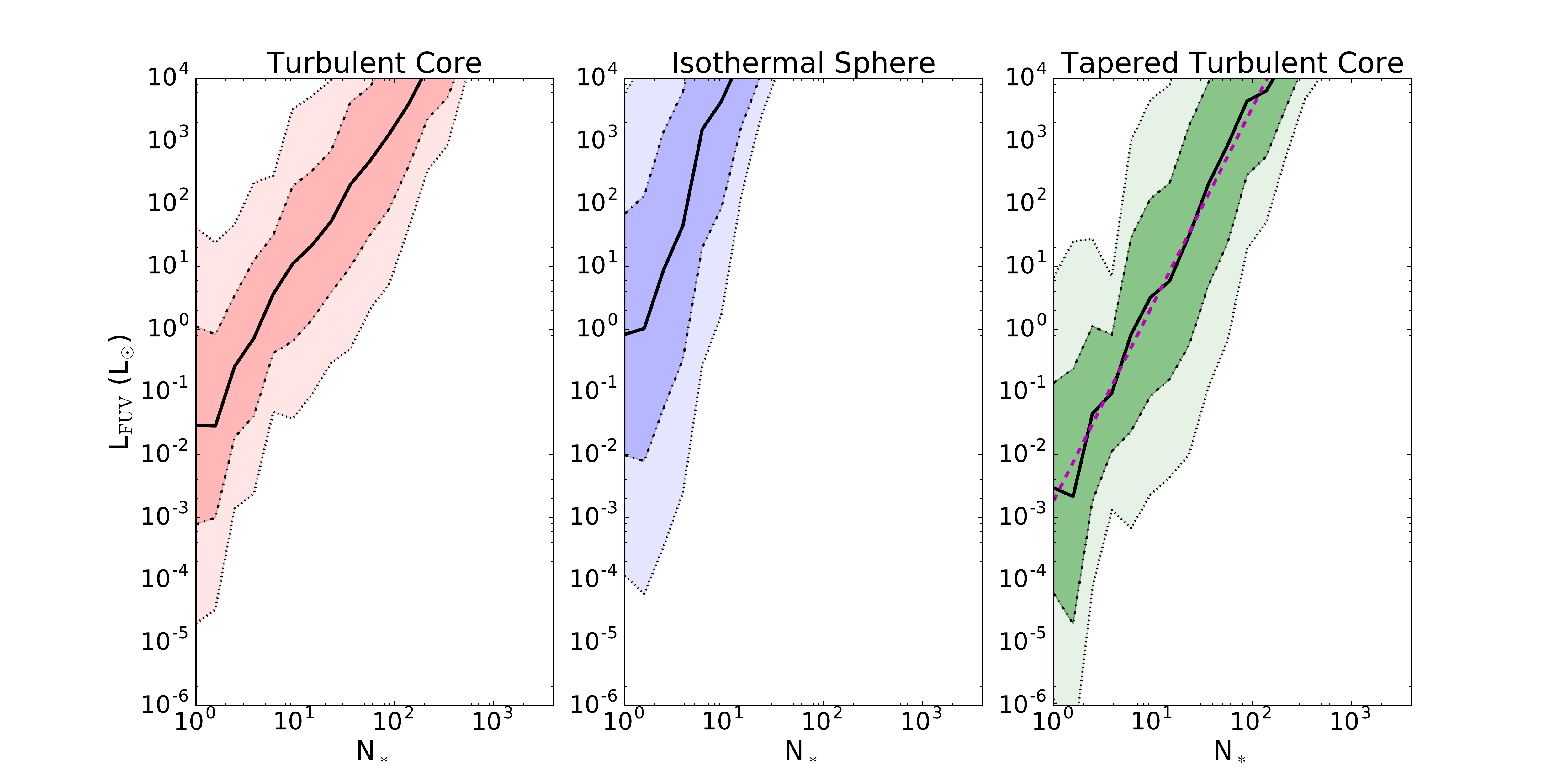}
\caption{\label{fig:clusterF} Cluster FUV luminosity versus the number of stars  for three different accretion histories. The black solid lines indicate the mean of the cluster distributions. The dark and light colored bands indicate the 1 and 2 $\sigma$ spread in the cluster luminosity. The magenta dotted line is the best fit for the TTC model (Equation \ref{eq:LFUV})}
\end{figure*}

Figure \ref{fig:clusterF} shows the PLF model predictions for the total FUV luminosity as a function of cluster size. The TC and TTC models exhibit significant spread in the predicted FUV for modest cluster sizes due to stochastic sampling of intermediate and high-mass stars, which contribute most of the FUV radiation. The IS PMF is narrower, which produces slightly better sampling. The spread is magnified in the TC and TTC models, because they assume a broad range of accretion rates. At large cluster masses the luminosity spread diminishes for all three models.  All accretion histories show a super-linear trend for small clusters, with the TTC model exhibiting the steepest dependence on cluster size: 
\begin{equation}\label{eq:LFUV}
\log L_{\rm FUV} = 
\begin{cases}
3.13 \cdot\log N_* - 2.73 & \text{$\log N_* < 2.42$} \\
\log N_* + 4.84 & \text{$\log N_* \ge 2.42$}
\end{cases}
\end{equation}

\begin{figure*}
\centering
\includegraphics[width=\textwidth]{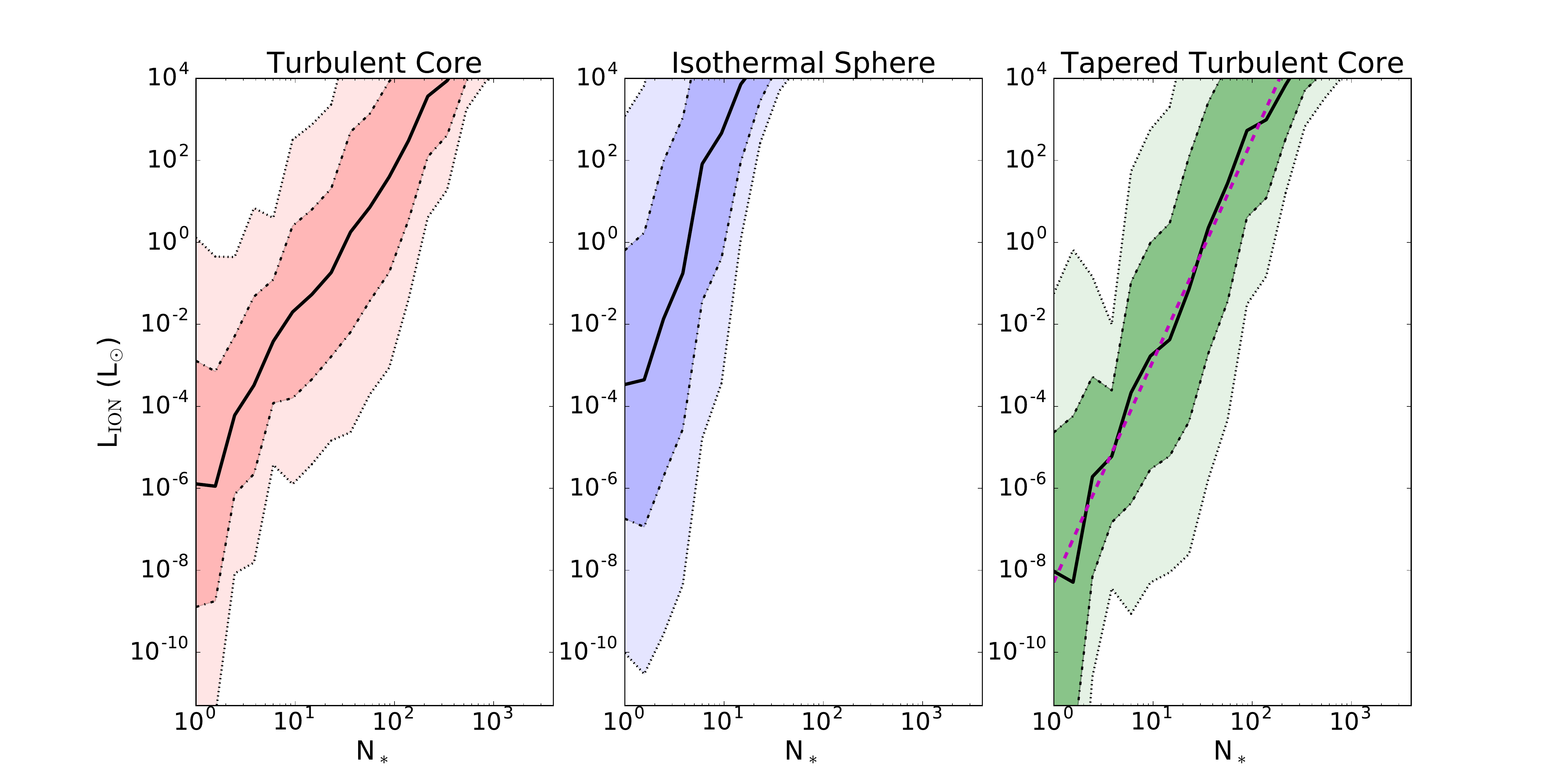}
\caption{\label{fig:clusterI} Cluster ionizing luminosity versus the number of stars  for three different accretion histories. The black solid lines indicate the mean of the cluster distributions. The dark and light colored bands indicate the 1 and 2 $\sigma$ spread of the distribution. The best linear fit to the means of the cluster distributions are annotated on the plot. The magenta dotted line is the best fit for the TTC model (Equation \ref{eq:LION})}
\end{figure*}

Figure \ref{fig:clusterI} shows the total ionizing luminosity, which exhibits a similar trend to the FUV component. Stochastic sampling of the highest mass protostars (future O and B stars),  which are the source of all ionizing radiation, creates larger scatter in the models.  Because O stars dominate the budget of ionizing radiation,  clusters with $N_*<10^4$,  which do not perfectly sample the high-mass end of the PMF, continue to  exhibit a large amount of statistical variation. The steeper slope is due to the strong dependence of  accretion rate on stellar mass and the higher peak accretion rates. The TTC model again exhibits the steepest dependence on cluster size:
\begin{equation}\label{eq:LION}
\log L_{\rm ION} = 
\begin{cases}
5.4 \cdot\log N_* - 8.29 & \text{$\log N_* < 2.42$} \\
\log N_* + 4.78 & \text{$\log N_* \ge 2.42$}
\end{cases}
\end{equation}

Overall, the models predict that once star formation commences a substantial amount of FUV radiation permeates the natal cloud. For lower-mass clusters the predicted amount of ionizion is very small, while a substantial amount of ionizing luminosity is expected in the highest mass clusters, such as the ONC complex or Cygnus X. In all cases, {\it statistical sampling introduces significant variation, which could drive environmental differences in clouds forming clusters with similar sizes.} 

\subsection{Cloud Properties and Abundances}

\begin{figure*}
\plotone{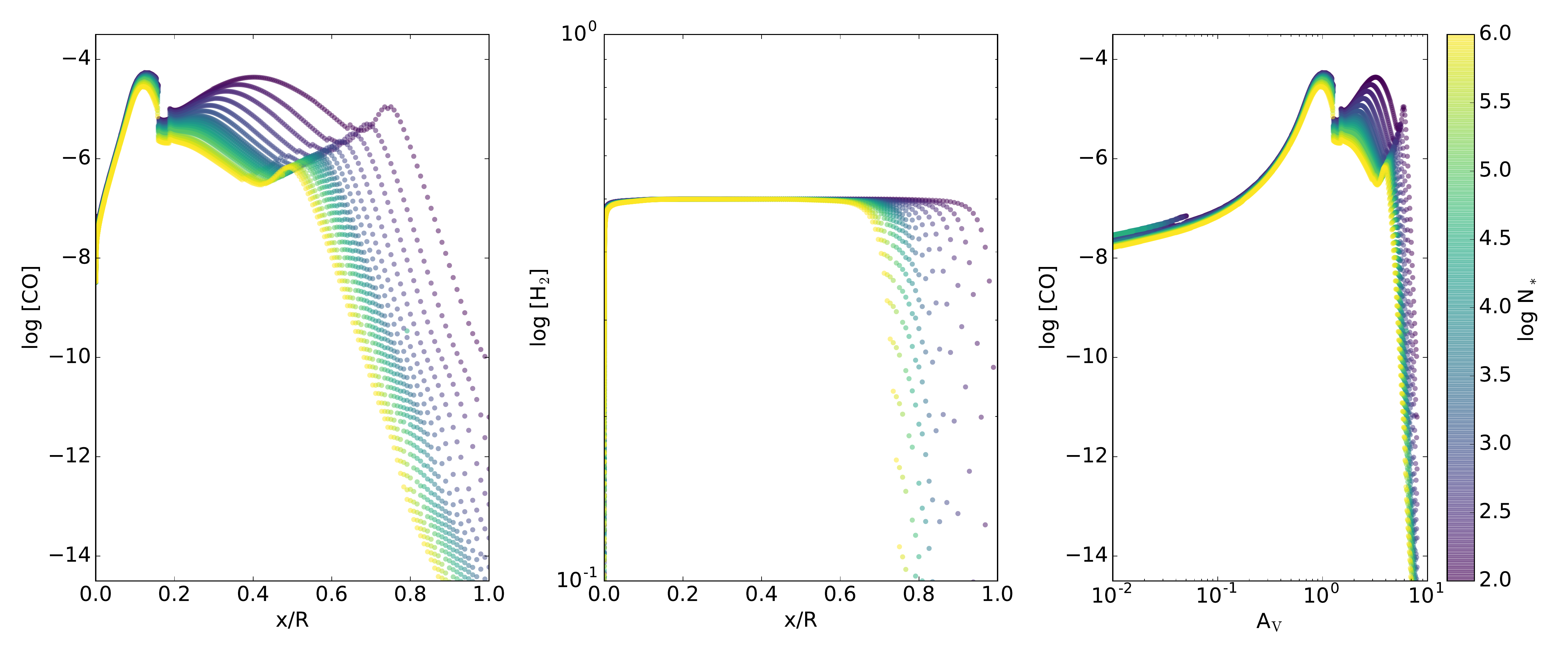}
\caption{\label{fig:cmchem1}Left: Fractional abundance of CO versus distance into the cloud with R = 4.1 pc for model CM\_1000D\_1kms\_1$\xi$. The coordinate, $x$, is measured such that $x=0$ at the cloud surface. Middle: Fractional abundance of H$_2$ as a function of distance into the cloud. Right: Fractional abundance of CO versus A$_V$. The color indicates the number of stars in the cluster, where purple corresponds to 100 stars and yellow corresponds to 10$^6$ stars.}
\end{figure*}

In this section, we use Model CM\_1000D\_1kms\_1$\xi$ to study the effects
of internal embedded sources on the chemical distribution within a cloud. Figure \ref{fig:cmchem1} shows the abundance of H$_2$ and CO as a function of cloud depth and extinction (A$_V$), where $x/R=0$ is the surface. At low A$_V$ the models for all cluster sizes are similar since the chemistry is dominated by the external radiation field.  The H$_2$ abundances converge to $\sim$ 0.5, which indicates that nearly all the H is in H$_2$. 

The embedded FUV sources ($x/R=1$) create a shell of H$_2$, which becomes progressively thinner with increasing cluster size. For $N_*=10^6$, the H$_2$ shell is only $\sim$60\% of the total cloud radius. In addition, the amount of CO is reduced even in the region that remains molecular. This is because the column density of material that provides self-shielding is much lower. Consequently, the embedded sources significantly alter the CO abundance profile compared to the typical 1D PDR model. Without embedded sources, the CO abundance asymptotically approaches a value around $10^{-4}$ at high A$_V$. However, the model predicts that CO is effectively dissociated by $A_V \geq 7$ for all clusters. Increasing the cluster size from $N_*\sim 100-10^6$ causes 2 orders of magnitude difference in the CO abundance at $A_V = 4$.

\begin{figure*}
\plotone{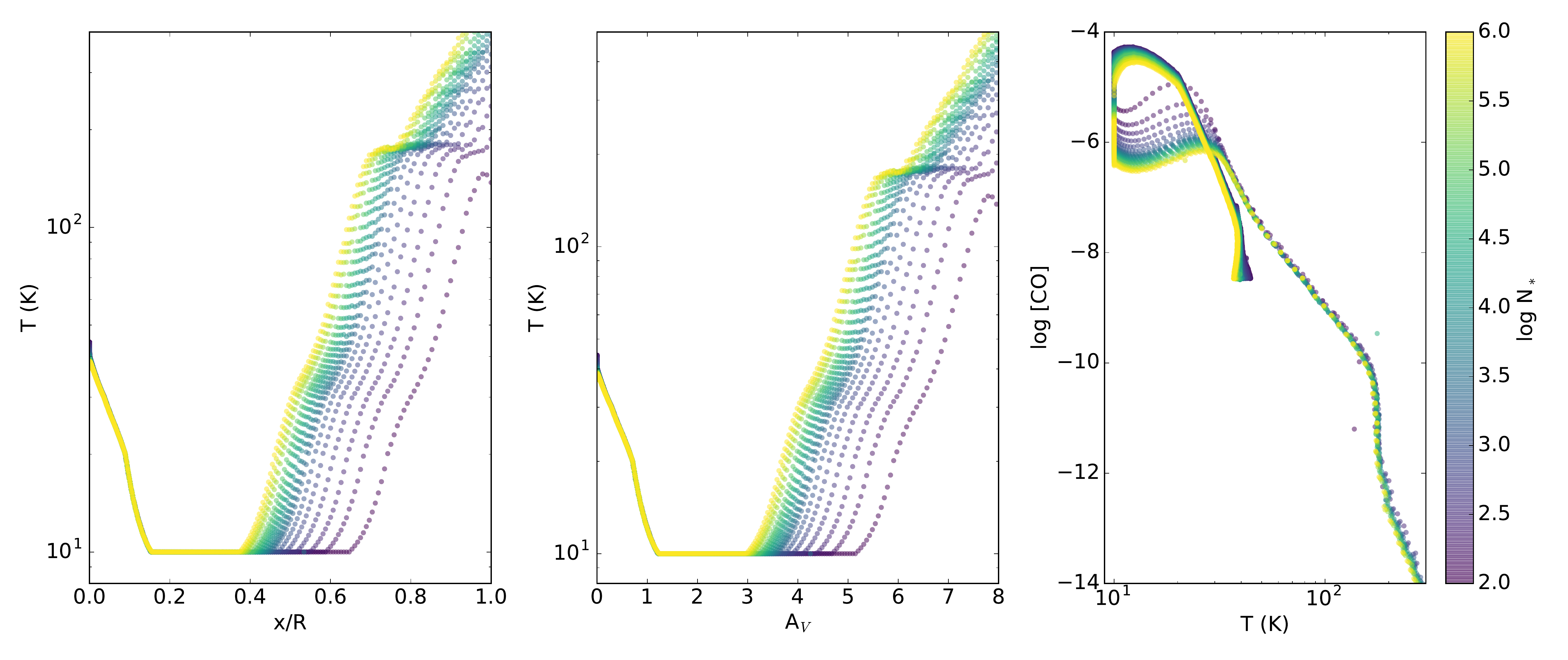}
\caption{\label{fig:cmchem2} Left: Temperature as a function of distance with R = 4.1 pc for model CM\_1000D\_1kms\_1$\xi$.  The coordinate, $x$, is measured such that $x=0$ at the cloud surface. Middle: Temperature as a function of A$_V$. Right: Phase plot showing the fractional abundance of CO versus gas temperature. The color indicates the number of stars in the cluster, where purple corresponds to 100 stars and yellow corresponds to 10$^6$ stars.}
\end{figure*}

Figure \ref{fig:cmchem2} shows the temperature structure of the cloud. {\sc 3d-pdr} determines the temperature by balancing the heating and cooling as described above. Without embedded sources, the cloud cools to a temperature of 10 K when $A_V \geq 1$. The model results show that the embedded sources have a strong impact, heating the high $A_V$ gas to hundreds of Kelvin. Comparing this temperature structure to the abundance profiles in Figure \ref{fig:cmchem1} indicates there is a large amount of warm CO. These temperatures lead to higher excitation,  so more emission preferentially comes from higher rotational levels.

The far right panel of Figure \ref{fig:cmchem2} displays
the CO abundance as a function of gas temperature. This phase digram shows a tight correlation between gas hotter than approximately 50 K and decreasing CO abundance. At cold temperatures, the phase diagram is more complicated due to the formation and destruction of CO at various points in the cloud.

\subsection{Variation of X$_{\rm CO}$}
In this section we investigate how changes in chemistry due to the presence of embedded sources impact the observed CO emission. We control for other factors including the cluster size, star formation efficiency, turbulent linewidth and gas density.

\subsubsection{Cluster Size with Fixed Cloud Mass}

\begin{figure}
\plotone{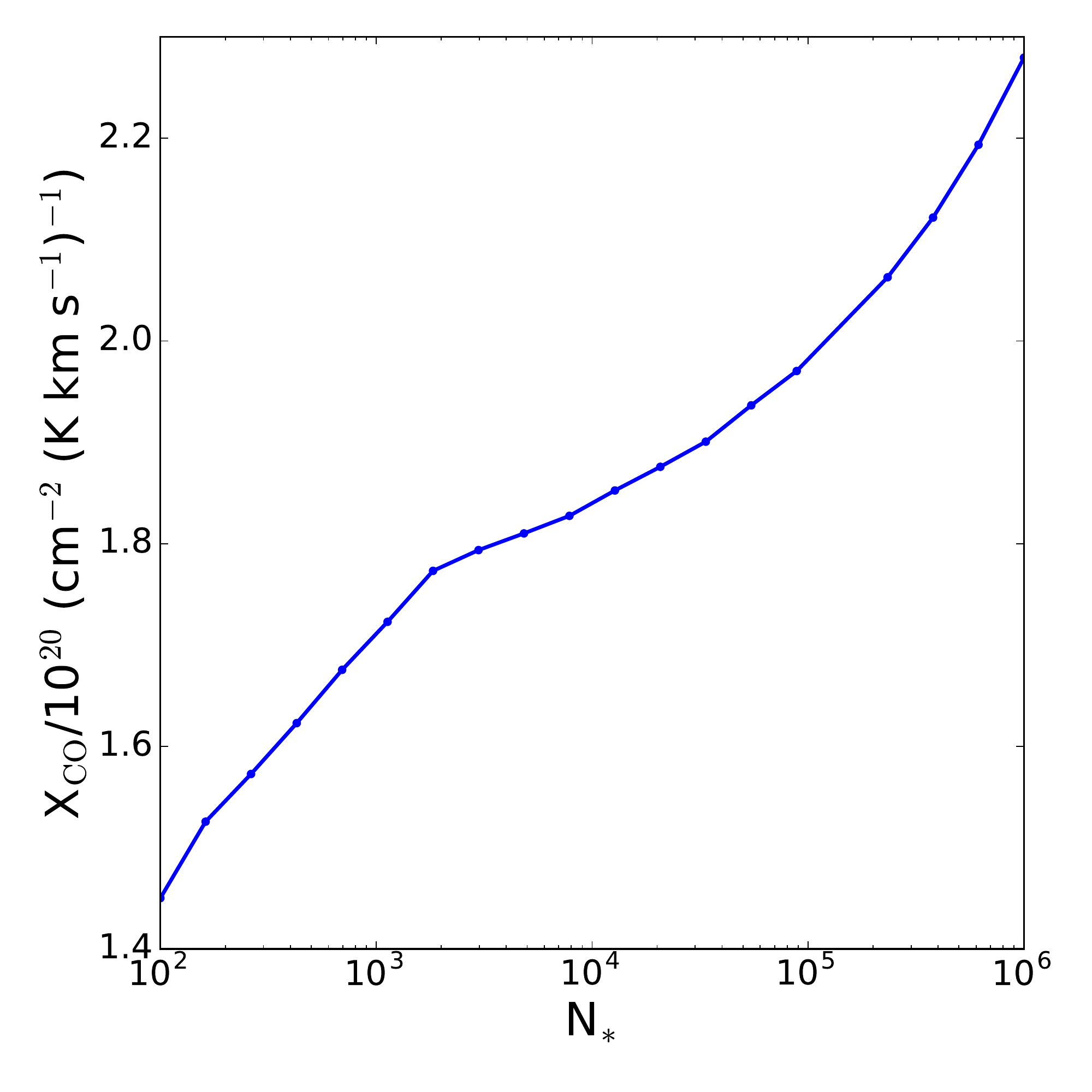}
\caption{\label{fig:cmxfac}X$_{\rm CO}$ normalized by 10$^{20}$ as a function of the number of stars, N$_*$, for model CM\_1000D\_1kms\_1$\xi$.}
\end{figure}

We first consider the simplest cloud model,  Model CM\_1000D\_1kms\_1$\xi$, which holds the cloud mass fixed for all cluster sizes. Figure \ref{fig:cmxfac} shows the model predictions for X$_{\rm CO}$ as a function of the number of stars in the cluster. X$_{\rm CO}$ approaches $10^{20}$ cm$^{-2}$ (K km s$^{-1}$)$^{-1}$ for small cluster sizes but increases steeply for large clusters. The abundance and temperature profiles in Figures \ref{fig:cmchem1} and \ref{fig:cmchem2} show the cause of the increase. For a large number of stars, the amount of FUV radiation increases super-linearly reducing the shell of molecular gas and the column density of H$_2$. However, due to the high optical depth of the CO (1-0) line the intensity is dominated by emission near the surface of the cloud. While more CO is dissociated due to the embedded FUV radiation, the gas also exhibits higher temperatures. These competing factors cancel, producing only a factor 2 change in X$_{\rm CO}$ over four dex of $N_*$. This insensitivity to cluster size is encouraging, since it seems to suggest that X$_{\rm CO}$ is largely invariant. However, our model assumptions break down for large clusters when the stellar mass becomes much greater than the gas mass.

\subsubsection{Cluster Size with Varying Cloud Mass}

\begin{figure}
\plotone{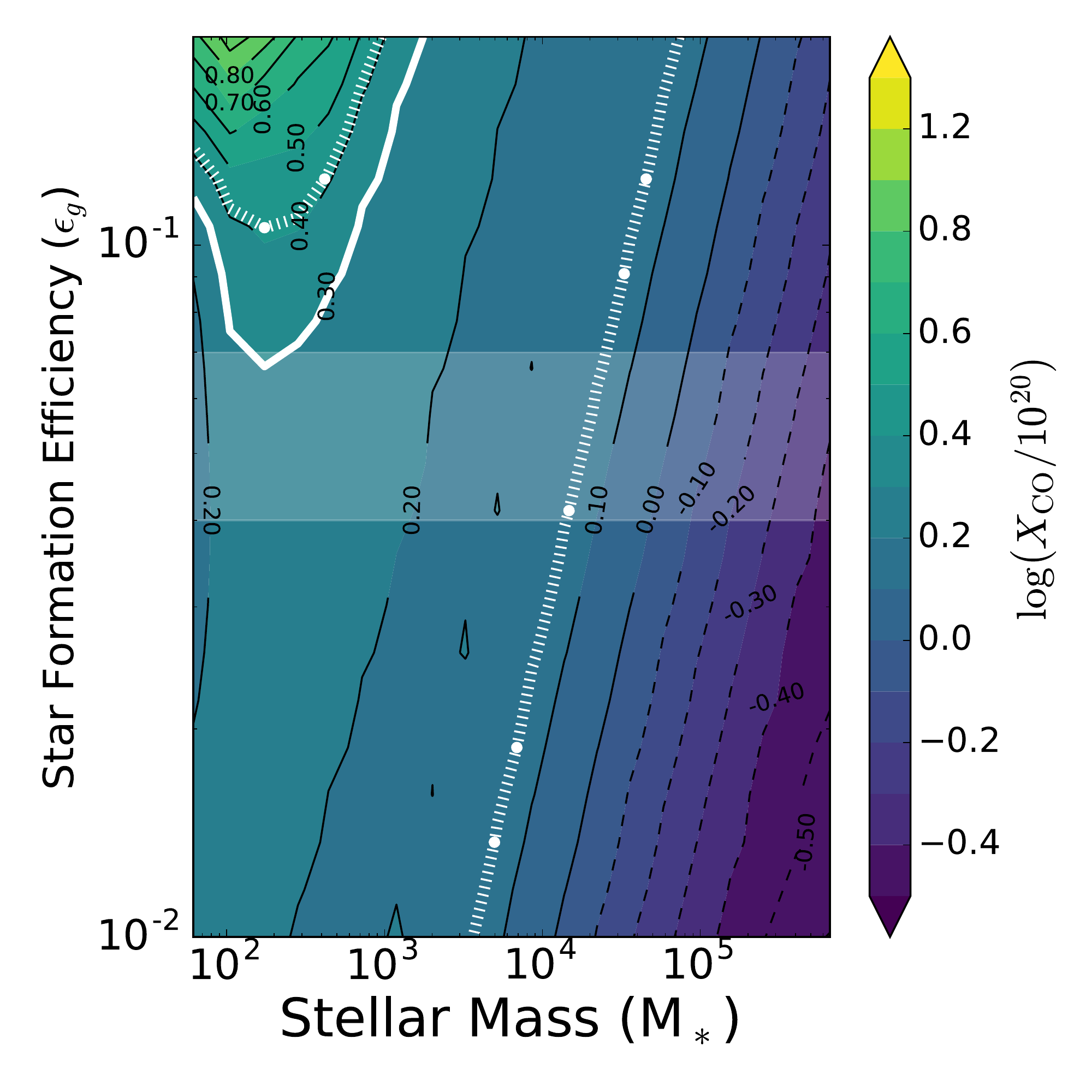}
\caption{\label{fig:meffxfac} Contour plot of X$_{\rm CO}$ as a function of the final star cluster mass, $M_*$, and the efficiency, $\varepsilon_g$, for model CE\_1000D\_V\_1$\xi$. The color scale indicates the logarithm of X$_{\rm CO}$  normalized by 10$^{20}$. The white solid contour is the typical Milky Way X$_{\rm CO}$ and the white dotted contours are the 30\% error bars \citep{2013ARA&A..51..207B}. The horizontal white band marks the star formation efficiency estimated for local Gould Belt clouds \citep{2013AJ....145...94D}.}
\end{figure}

Figure \ref{fig:meffxfac} shows X$_{\rm CO}$ as a function of cluster mass, $M_*$, and the star formation efficiency, $\varepsilon_g$. Here, $M_*$ is the total protostellar mass ($\Sigma_i ^{N_*} m_i$). This figure shows the opposite trend to the constant mass model shown in Figure \ref{fig:cmxfac}. For fixed values of the efficiency, X$_{\rm CO}$ drops by a factor of a few as the cluster mass increases by 4 dex. This mainly occurs as a result of assuming the molecular cloud is virialized. The corresponding larger linewidths increase the integrated CO intensity causing  X$_{\rm CO}$ to decline. This model is more physically motivated than the simpler constant mass model; however, it shows that the gas-to-star conversion has a significant impact on the relationship between the column density and CO emission.

In Figures \ref{fig:meffxfac}-\ref{fig:meffxfactnoFUV}, the solid white contour indicates the average X$_{\rm CO}$ measured in the MW, X$_{\rm CO}$ = $2\times 10^{20}$ cm$^{-2}$ (K km s$^{-1}$)$^{-1}$, and the dotted white contours indicate the $\pm$ 30\% error \citep{2013ARA&A..51..207B}. Our predicted X$_{\rm CO}$ are consistent with the measured MW values for a large fraction of the parameter space. If we further constrain to look at the region of parameter space encompassing measured star formation efficiencies (See below), the model is consistent for clusters between $N_* \approx$ 20 - 10$^4$. The protostar surveys, mentioned above, span that range of cluster sizes for local star forming regions where X$_{\rm CO}$ measurements are best measured.

\subsubsection{Star Formation Efficiency}

An important consideration is the relative amount of mass in stars and gas as codified by the star formation efficiency, $\epsilon_g$. Figure \ref{fig:meffxfac} shows $X_{\rm CO}$ increases with $\epsilon_g$ for fixed cluster mass. For large clusters, X$_{\rm CO}(\epsilon_g)$ a factor of two difference over 1.5 dex of star formation efficiency. These clusters have gas masses sufficient for their optical depths to minimize the impact of the embedded feedback. Therefore, the CO line emission is not much affected by radiation feedback from the embedded cluster. The change for the largest clusters is due to the change in velocity dispersion. For smallest clusters, the change in X$_{\rm CO}$ with $\epsilon_g$ is a factor of four. For the smallest clusters, the increased sensitivity to the embedded clusters is due to the reduction in cloud optical depth. The trend here is driven by irradiation by 
the embedded clusters.

The white band in Figure \ref{fig:meffxfac} shows the measured star formation efficiencies from the \citet{2013AJ....145...94D} survey of Gould Belt clouds. Within the band, a significant amount of the parameter space is consistent with the local Milky Way average X$_{\rm CO}$ (in white contours). Furthermore,   X$_{\rm CO}$ is nearly constant for moderate cluster sizes, so our model  predicts the Milky Way average is representative of local molecular clouds. The model also predicts $X_{\rm CO}$ decreases by a factor of 5-10 in the largest clusters due to the increase in turbulent linewidth. 

\subsubsection{Mean Gas Density}

\begin{figure}
\plotone{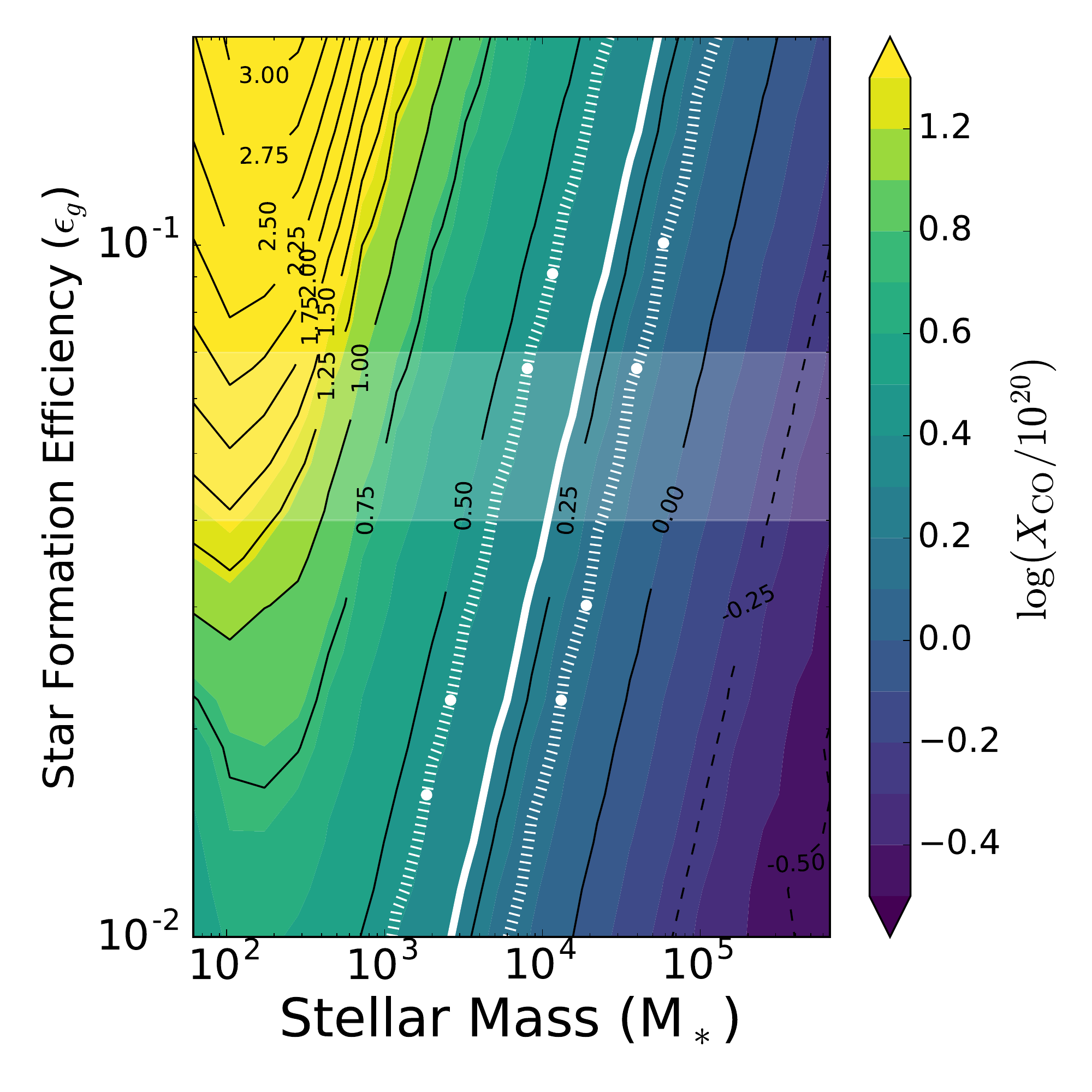}
\caption{\label{fig:meffxfacn500} Same as Figure \ref{fig:meffxfac} but for model CE\_500D\_V\_1$\xi$.}
\end{figure}

Molecular clouds have a range of mean densities. In this section, we explore the impact of the mean gas density on X$_{\rm CO}$. Model CE\_500D\_V\_1$\xi$ is the same model as the fiducial expect with n$_{\rm H}$ = 500 cm$^{-3}$. Figure \ref{fig:meffxfacn500} shows the same parameter space as the fiducial model shown in Figure \ref{fig:meffxfac}. The lower density causes X$_{\rm CO}$ to increase over much of the parameter space. Lowering the density also reduces the column density (and thus the dust extinction) making photochemistry more important. However,  changes in the amount of molecular hydrogen and the CO (1-0) emission compete and partially cancel. If there is a reduction in both, X$_{\rm CO}$ may increase but not by a large factor. In Figure \ref{fig:meffxfacn500} the overall trend remains the same as the fiducial model but is amplified for moderate and smaller clusters. There is no change for the largest clusters since they have sufficient mass such that changes in the interior abundances occur after the line has become opaque.

X$_{\rm CO}$ for small clusters is greatly amplified due to the lower dust extinction within the cloud. These clouds have significantly less CO emission compared to their H$_2$ column density, i.e., they have a larger fraction of ``CO-dark'' gas \citep[i.e.][]{2010ApJ...716.1191W}. Typically,  CO-dark clouds are assumed to have faint CO emission due to their low densities, but the gas here is CO deficient due to dissociation caused by the embedded sources. Consequently, the MW average values occupy only a narrow band across the parameter space. Within the range of typical star formation efficiencies, X$_{\rm CO}$ is only consistent for clusters with masses between 10$^3$ - 10$^4$ M$_{\odot}$. 

Prior work has also found that X$_{\rm CO}$ is sensitive to the gas density \citep{2006MNRAS.371.1865B, 2011MNRAS.412.1686S}. Densities higher than 10$^3$ cm$^{-3}$ make dust extinction more efficient, while lower densities enhance the effect of embedded clusters since more of the cloud is influenced by photochemistry. Our models predict X$_{\rm CO}$ is most sensitive to density for clouds forming small clusters with high star formation efficiencies. However, this trend may not be evident in observations since diffuse clouds are less likely to form stars with high efficiencies. 

\subsubsection{Turbulent Velocity Dispersion}

The turbulent velocity dispersion is an important factor in the calculated CO emission due to its influence on the line width and, hence, the optical depth. The line optical depth for a given line of sight is inversely proportional to the velocity dispersion. There is ample evidence that higher mass clouds have greater velocity dispersions and that many clouds are close to virial equilibrium \citep{2015ARA&A..53..583H}. Although a constant linewidth model is unphysical, it is useful to examine the importance of velocity information.
In this section, we study the effects of the turbulent velocity dispersion by comparing model CE\_1000D\_V\_1$\xi$ to model CE\_1000D\_1kms\_1$\xi$. 

\begin{figure}
\plotone{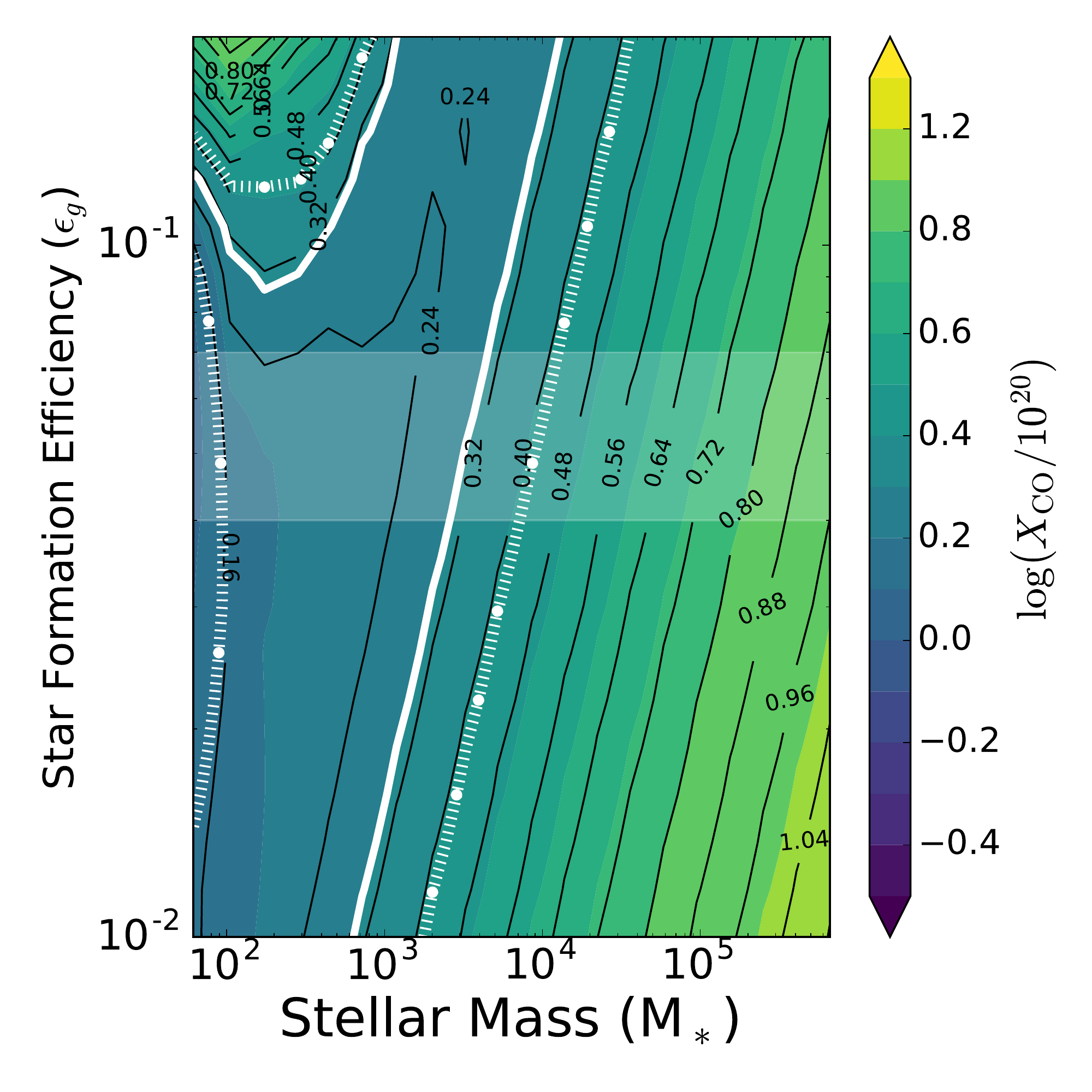}
\caption{\label{fig:meffxfact1kms} Same as Figure \ref{fig:meffxfac} for Model CE\_1000D\_1kms\_1$\xi$.}
\end{figure}

Figure \ref{fig:meffxfact1kms} shows X$_{\rm CO}$ across the parameter space assuming a constant turbulent linewidth of 1 km/s. X$_{\rm CO}$ exhibits a similar trend to that of the constant mass model shown in Figure \ref{fig:cmxfac}. A smaller turbulent velocity dispersion increases the line optical depth, which decreases the overall integrated line flux. The decline in flux, for the same H$_2$ distribution, increases X$_{\rm CO}$. This completely reverses the trend illustrated in Figure \ref{fig:meffxfac}. Thus, increasing velocity dispersion accounts for much of the decline in X$_{\rm CO}$ with increasing cloud mass, and the local velocity dispersion is essential to understanding trends in X$_{\rm CO}$.

\subsubsection{Cosmic Ray Ionization Rate}
We investigate the effect of the cosmic ionization rate, $\xi$,  which prior work indicates strongly influences $X_{\rm CO}$ \citep[e.g.,][]{2015ApJ...803...37B}. X$_{\rm CO}$ increases with $\xi$ due to the increased destruction of CO and overall decline in the emission. Higher cosmic ray fluxes also lead to higher gas temperatures, which in principle could cause X$_{\rm CO}$ to decline. However, a value of $\xi = 100$ is not high enough to make cosmic ray heating the dominant heating mechanism throughout the whole cloud \citep{2006MNRAS.371.1865B}. Model CE\_1000D\_V\_100$\xi$ adopts a cosmic ionization rate enhanced by a factor of 100 compared to the other models. An increase in the cosmic ray ionization rate is observed in environments with more star formation, such as those in ULIRGS \citep{2010ApJ...720..226P} and towards the central molecular zone of the Milky Way \citep{2016A&A...585A.105L}.

\begin{figure}
\plotone{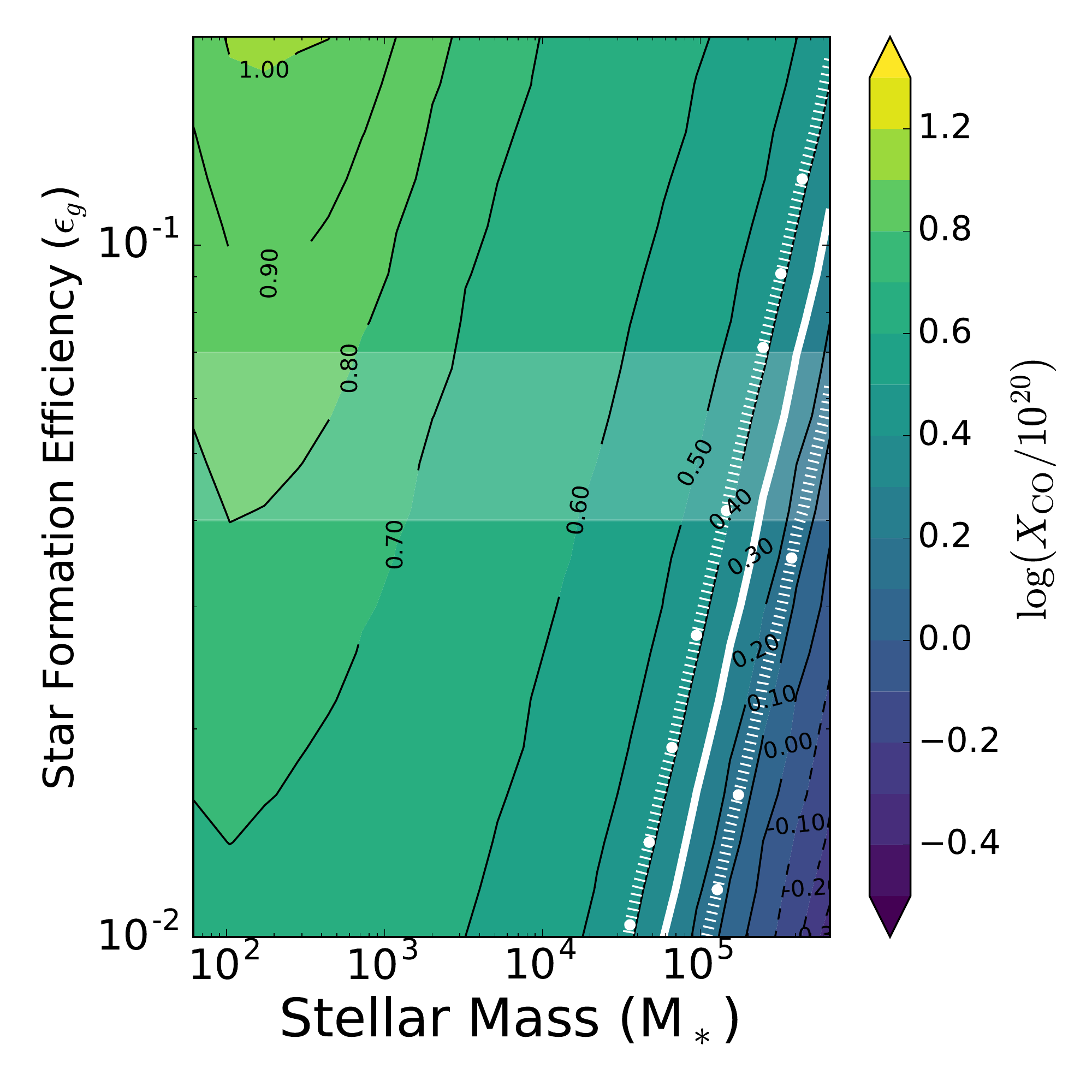}
\caption{\label{fig:meffxfactxi100} Same as Figure \ref{fig:meffxfac} for Model CE\_1000D\_V\_100$\xi$.}
\end{figure}

Figure \ref{fig:meffxfactxi100} shows X$_{\rm CO}$ for the enhanced cosmic ray ionization rate. The higher rate increases X$_{\rm CO}$ by a nearly constant value for all stellar masses. However, the overall trend remains, and the total spread is similar. Since X$_{\rm CO}$ increases, the fraction of the parameter space consistent with the measured Milky Way values declines.

\subsubsection{Impact of Internal Sources}
\begin{figure}
\plotone{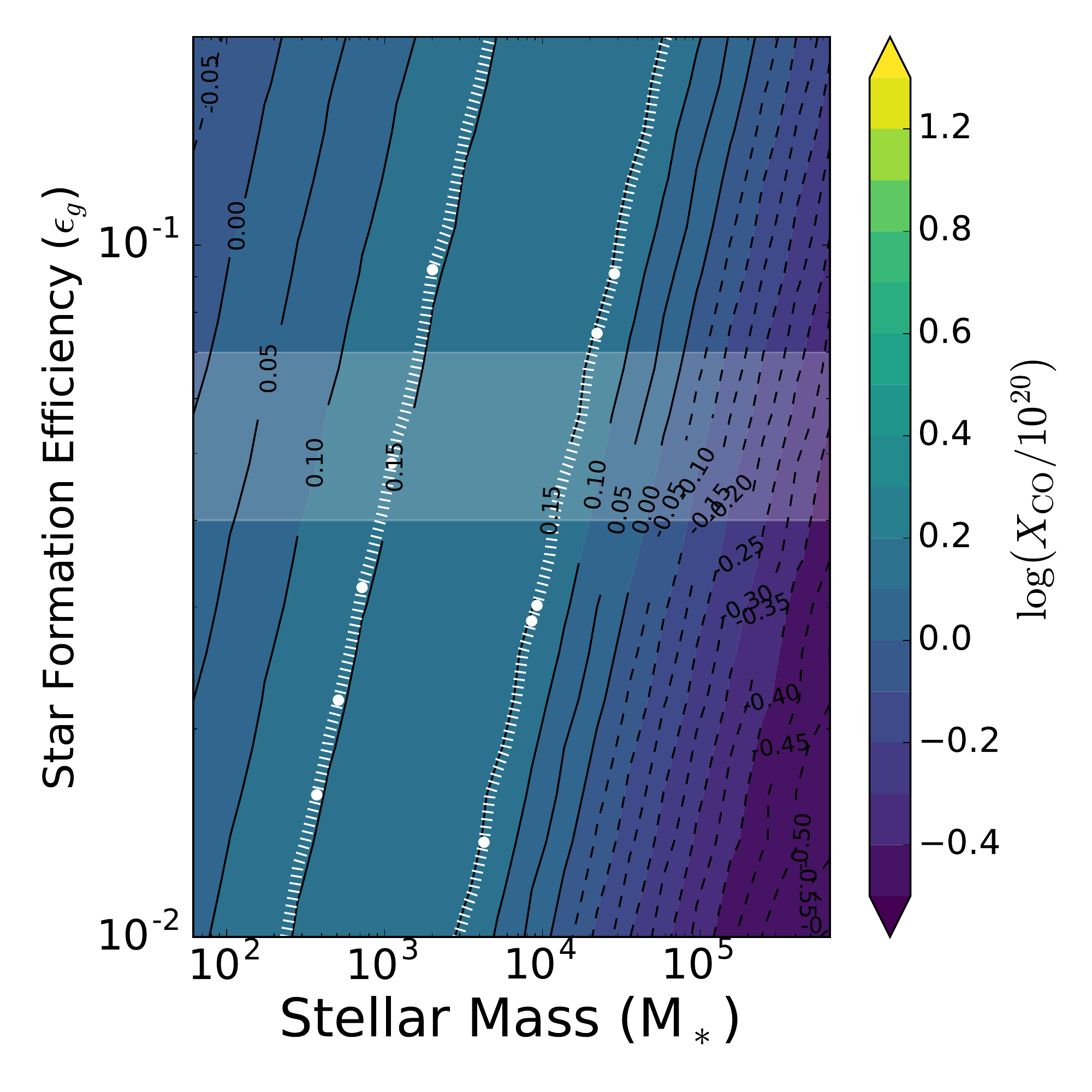}
\caption{\label{fig:meffxfactnoFUV} Same as Figure \ref{fig:meffxfac} except the internal FUV flux is not included in the chemistry modeling.}
\end{figure}

To constrain the impact of embedded sources, specifically, on $X_{\rm CO}$, model CE\_1000D\_V\_1$\xi$\_NS excludes the star cluster FUV. Figure \ref{fig:meffxfactnoFUV} shows Model CE\_1000D\_V\_1$\xi$\_NS with an external field only. Clusters with a mass greater than a few thousand solar masses show almost no difference in X$_{\rm CO}$ compared to the fiducial model with the inclusion on internal fields. Towards smaller clusters, the model without internal radiation shows an opposite trend. Without the internal FUV radiation, X$_{\rm CO}$ decreases towards small efficient clusters. Furthermore, X$_{\rm CO}$ decreases enough that the average MW value is no longer represented in the parameter space. The model values of X$_{\rm CO}$ within the local star formation efficiency band are only consistent with the lowest measured values.

The inclusion of internal FUV radiation increases X$_{\rm CO}$ for clusters within the sizes indicated in Figure \ref{fig:clusterL} towards MW average values. Large clusters are relatively unaffected because the turbulent linewidth dominates over chemical effects. Embedded photochemistry only affects the smaller clusters since CO emission is dominated by flux emitted closer to the surface. 

\begin{figure}
\plotone{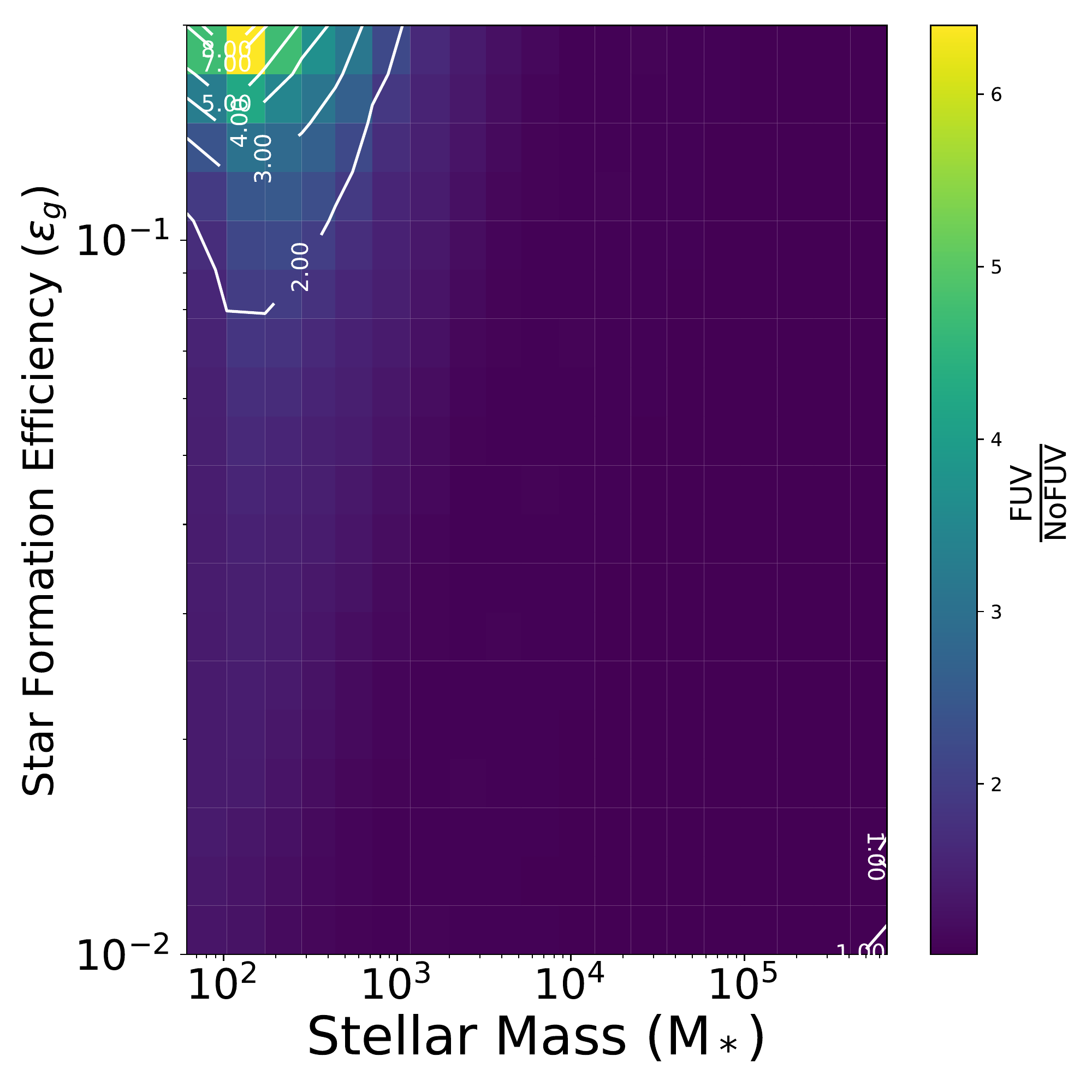}
\caption{\label{fig:xcodiff}Ratio of X$_{\rm CO}$ calculated with the embedded protostellar FUV (model CE\_1000D\_V\_1$\xi$) to X$_{\rm CO}$ calculated without (model CE\_1000D\_V\_1$\xi$\_NS). }
\end{figure}

Figure \ref{fig:xcodiff} shows the linear ratio of X$_{\rm CO}$ with embedded sources and without them. For most the parameter space, the embedded sources increase $X_{\rm CO}$ by 30-50\%. For small efficient clusters, the change is up to a factor of 8, increasing rapidly towards clouds with smaller gas mass. For these clouds, X$_{\rm CO}$ would likely be time-dependent, evolving with the protostellar population.

\section{Discussion}\label{sec:dis}

\subsection{Implications for Unresolved Star-Formation in Extragalactic Sources}

Measuring molecular gas mass in extragalactic sources relies on two dominant methods: dust observations in the infrared and sub-millimeter and CO emission. In the later case, the common procedure is to use some approximate conversion factor to calculate the total molecular gas mass within a galaxy. Molecular gas measurements for local galaxies have resolutions of tens to hundreds of parsecs \citep[e.g.][]{2017MNRAS.471.2917K, 2017A&A...601A.146C}. Furthermore, many of the galaxies targeted are actively star-forming. Since the measured CO integrated flux is an average over the spatially larger star-forming regions, our results suggest embedded star formation must be taken into account. 

Our model results show that for the largest clusters, there is little impact from the embedded FUV radiation and the conversion factor is instead dominated by the turbulent line width. However, for smaller clusters in the range of hundreds to thousands of stars, the embedded radiation has a clear effect. These smaller clusters have X$_{\rm CO}$ values factors of 3-10 larger than otherwise assumed without the embedded clusters. Furthermore, excluding embedded FUV sources will bias chemical models towards either lower densities, higher external radiation or higher cosmic ray fluxes. Our X$_{\rm CO}$ factors presented here are lower limits since our models are one-dimensional constant density slabs. Real clouds have significant structure and porosity, and the embedded protostars are not tightly grouped into a central cluster but distributed throughout the cloud. Both of these effects would serve to increase the embedded FUV throughout the cloud, amplifying these trends.

Carbon monoxide has been measured in galaxies out to high redshifts using large single-dish integrated line measurements \citep[e.g.,][]{2015MNRAS.454.3485Y}. At these redshifts, the star formation rate densities are typically much greater than present-day values \citep{2014ARA&A..52..415M}. These measurements are dominated by the brightest CO regions, which we predict inherently correspond to lower values of X$_{\rm CO}$. 

\subsection{Implications for Dense Gas Tracers of Star Formation}

Many molecular gas surveys use dense gas tracers to more directly measure the molecular gas undergoing star formation. Tracers such as HCN and HCO$^+$ are the most common alternatives to CO due to their relatively high abundances \citep[e.g.][]{2017ApJ...836..101C, 2007ApJ...660L..93G}. Ammonia (NH$_3$) is also readily observed in local galaxies \citep[e.g.][]{2011A&A...534A..56L}, even though it is associated with dense gas and has a low abundance, because it has a low critical density. Optically thin isotopologues of CO such as $^{13}$CO and C$^{18}$O are often used for line ratio diagnostics \citep{2013ARA&A..51..207B}. Because they are optically thin, emission from these tracers is sensitive to the conditions of the high A$_V$ gas, especially molecules such as NH$_3$ and HCN. Strong FUV radiation from embedded forming star clusters not only dissociates the molecules but heats the gas in the vicinity to hundreds of degrees. Therefore, our work underscores the importance of considering the embedded FUV when modeling optically thin emission from regions expected to have accreting protostars. 

In some ways, this is not a novel conclusion. A variety of prior observational work has studied the evolution of gas chemistry near protostars XX, and observations of high-mass protostars, in particular show significant chemical time variation with protostellar evolution \citep[i.e.,][]{1996ApJ...460..359H, 2014prpl.conf..243K}. Our results build on the previously acknowledged importance of protostellar feedback to provide a framework for quantifying the  impact of feedback on chemistry at cloud scales in addition to the well-studied smaller scales of individual protostars.

\subsection{X$_{\rm CO}$ Variation within Galaxies}

A large number of surveys have studied X$_{\rm CO}$ variation within the MW. For example, \cite{1995ApJ...452..262S} and \cite{2004A&A...422L..47S} investigate the radial dependences of X$_{\rm CO}$. Both results, although using different methods, conclude that X$_{\rm CO}$ increases with radius. There are various possible explanations for this trend, Of particular import is the role of the turbulent linewidth and the mass surface density of clouds. In the center of the galaxy molecular clouds not only have larger masses on average but also larger column densities compared to clouds in the outer regions of the galaxy \citep{2010ApJ...723..492R}. Furthermore, the overall galactic mass surface density decreases with radius in the MW except for a slight increase around 4 kpc. The star formation rate (SFR) surface density also generally decreases except for the same bump at 4 kpc \citep{2012ARA&A..50..531K}. \cite{1995ApJ...452..262S} measure X$_{\rm CO}$ in the center of the MW to be $\log \frac{{\rm X_{\rm CO}}}{10^{20}} \approx -0.5$. Our work produces this value for large inefficient star-forming molecular clouds, especially those subject to a strong cosmic ray flux. The outer galactic values in \cite{1995ApJ...452..262S} are between $\frac{{\rm X_{\rm CO}}}{10^{20}} \approx 0.6 - 1.0$, which are represented in our parameter space by small to intermediate molecular clouds for a mean density of $10^3$ cm$^{-3}$ or smaller star-forming clouds with a mean density of $500$ cm$^{-3}$. 

Similar trends are observed in other nearby galaxies. \cite{2013ApJ...777....5S} measured X$_{\rm CO}$ using high resolution {\it Herschel} maps of 26 nearby disk galaxies. The survey showed that nearly all galaxies exhibit a decrease in X$_{\rm CO}$ in their inner regions. For example, NGC 6946 is a nearby disk galaxy around 7 Mpc away and one of the galaxies included in \cite{2013ApJ...777....5S}. NGC 6946 has also been shown to have a radially decreasing molecular gas and SFR surface density \citep{2012ARA&A..50..531K}. The model parameter space used in this work has a mass surface density that increases from the upper left corner to the lower right corner. Furthermore, \cite{2013ApJ...777....5S} finds the average central $\log \frac{{\rm X_{\rm CO}}}{10^{20}} \approx -0.5 - 0.2$ consistent with large star-forming molecular clouds in this work. We find that when embedded sources are included we replicate the trends between molecular gas surface density and X$_{\rm CO}$ in NGC 6947. We stress, however, that this trend does not appear without the FUV radiation from embedded star formation.

\subsection{Comparison to Other Astrochemistry Studies}

\cite{2006MNRAS.371.1865B} used the {\sc ucl-pdr} code\footnote{{\sc 3d-pdr} extends {\sc ucl-pdr} to 3D, so the underlying approaches are very similar.}  \citep{2005MNRAS.357..961B} to perform a parameter study of X$_{\rm CO}$ as a function of A$_V$. This cannot be directly related to a cloud integrated X$_{\rm CO}$, but rather to the cloud average A$_V$, but the trends are similar. \cite{2006MNRAS.371.1865B} show that in high-density environments X$_{\rm CO}$ is only weakly affected by the impinging UV field with X$_{\rm CO}$ decreasing slightly over 5 orders of magnitude of increasing field. The trend reverses at low A$_V$ where slight increases in the FUV field significantly increase X$_{\rm CO}$. In our study, the external field is fixed while the internal field is increased, and we find low $A_V$ gas is relatively unaffected by the embedded sources. 

\cite{2012MNRAS.421.3127N} studied the effect of galaxy mergers on X$_{\rm CO}$. They compared simulations of quiescent star forming discs with merging starburst systems and found that the local variation within the galaxy is a smooth function of metallicity. However, starburst systems, which have much higher SFRs, exhibit a lower X$_{\rm CO}$ factor. They attributed the lower value to an increase in temperature caused by heating from young high-mass stars and larger gas velocity dispersions. This is in good agreement with our model, where the inclusion of FUV radiation from embedded star formation systematically increases the temperature locally, while more massive, turbulent clouds have lower X$_{\rm CO}$.

Recently, \cite{2015MNRAS.452.2057C} performed simulations to study how X$_{\rm CO}$ varied with SFR. They fixed bulk properties such as the total mass and initial turbulent field  and  varied environmental factors that are thought to correlate with star formation rate. They linearly increased the external FUV field and cosmic ray ionization rate with the assumed star formation \citep{2010ApJ...720..226P}. They found X$_{\rm CO}$ increases with star formation rate, contrary to other studies. In fact, their models are similar to our constant mass model, CM\_1000D\_1kms\_1$\xi$, (Figure \ref{fig:cmxfac}) and constant velocity model CE\_1000D\_1ks\_1$\xi$ (Figure \ref{fig:meffxfact1kms}), in which  X$_{\rm CO}$ also increases with cluster mass. Here, this is due to the rapid photodissociation of CO, such that the clouds become CO-deficient or rather ``CO-faint''.  The constant mass model is also represented in our constant efficiency models by using a fixed cluster mass and increasing the efficiency. X$_{\rm CO}$ increases as a function of star formation efficiency in all of our CE models. Our results show the same qualitative trends as \cite{2015MNRAS.452.2057C} when considering models that keep bulk hydrodynamic properties fixed. Keeping the velocity dispersion constant as the star formation activity increases leads to an increasing X$_{\rm CO}$ as a function of star formation activity. 

Previous theoretical studies probed the star formation rate by changing the external environment. Higher SFR clouds are bathed in stronger FUV fields and in some cases \citep{2012MNRAS.426.2142L, 2015MNRAS.452.2057C} experience higher cosmic ray ionization. The cosmic-ray ionization rate also correlates with the supernova rate and thus the SFR. The result of the scaling between the supernova rate and the SFR creates the linear scalings, $\chi \sim \chi_0 \times {\rm SFR}$ and $\xi \sim \xi_0 \times {\rm SFR}$, where $\chi_0$ and $\xi_0$ are the MW ISRF and cosmic ray ionization rate. The scalings of the impinging FUV radiation and cosmic ray flux with the SFR apply for galactic-wide studies where a galaxy-averaged SFR is used. On smaller scales these correlations do not hold due to the star formation activity becoming more stochastic.

\section{Summary and Conclusions}

This paper presents an approach coupling a semi-analytic protostellar cluster model with a PDR code to study of the effects of FUV stellar feedback on the natal physical and chemical environment of molecular clouds. We create a semi-analytic model to calculate cluster luminosities as a function of the number of protostars. We calculate the total, far-ultraviolet and ionizing luminosities for three different accretion models: Isothermal Sphere (IS), Turbulent Core (TC) and Tapered Turbulent Core (TTC) using the Protostellar Luminosity Function (PLF) formalism. We compare the model predictions against observations of three different surveys \citep{2013AJ....145...94D, 2012AJ....144...31K, 2014AJ....148...11K} and find our results for the TTC model fit the observations well. We present fits to the model predictions for the different luminosities as a function of cluster size summarized below for the TTC model:
\begin{subequations}
  \begin{equation}
  \log L_{\rm Bol} = 1.96 \cdot\log N_* + 0.18 \quad\text{if $\log N_* < 2.78$} 
  \end{equation}

  \begin{equation}
  \log L_{\rm FUV} = 3.13 \cdot\log N_* - 2.73 \quad\text{if $\log N_* < 2.42$}
  \end{equation}

  \begin{equation}
  \log L_{\rm ION} = 5.4 \cdot\log N_* - 8.29  \quad\text{if $\log N_* < 2.42$}
  \end{equation}
\end{subequations}
with the equations becoming linear after the indicated break.

We use the photodissociation region (PDR) code {\sc 3d-pdr} to model the chemistry of molecular clouds hosting forming embedded star clusters assuming two different physical models: a constant mass model, where the cloud contains a fixed 10$^4$ M$_{\odot}$ of gas, and a constant efficiency model where the total gas mass scales with the cluster mass and the star formation efficiency parameter. Using the constant mass model, we study the chemical and physical effects of the embedded FUV radiation in detail. We find that the embedded FUV flux significantly increases the temperature of the high A$_V$ gas, raising the temperature to hundreds to thousands of degrees Kelvin deep within the cloud. Furthermore, we find that increasing the cluster mass creates a thinner shell of H$_2$ and CO, reducing the amount of CO by orders of magnitude, even at the A$_V = 1$ surface.

We calculate X$_{\rm CO}$ as a function of cluster mass for both physical models. The constant mass model, which also assumes a constant velocity dispersion, has an X$_{\rm CO}$ that increases with cluster mass. However, the increase is small: only a factor of 2 increase over four orders of magnitude in cluster mass. In contrast, the constant efficiency models show the opposite trend. In these models, the velocity dispersion is calculated assuming the cloud is in virial equilibrium. We find that X$_{\rm CO}$ decreases with higher cluster masses, although there is a slight increase for higher efficiencies due to  their lower column densities. Altogether, the trends over four orders of magnitude in cluster mass and two orders of magnitude in star formation efficiency amount to a 1.5 dex variation in X$_{\rm CO}$. Most of the parameter space is consistent with the measured MW values, and we find that including feedback from embedded clusters improves the agreement with observations.

We also investigate the effect of three different parameters on X$_{\rm CO}$  for the constant efficiency model. We calculate X$_{\rm CO}$ using mean gas densities of $n_{\rm H}$ = 500 cm$^{-3}$ and $n_{\rm H}$ = 1000 cm$^{-3}$. We find that the qualitative trend remains the same, although for the lower density cloud the dispersion is over three dex over the whole parameter space. Reducing the density increases the typical $X_{\rm CO}$, decreasing the agreement with the average MW value. We also fix the velocity dispersion at 1 km s$^{-1}$. In this case, the trend reverses. The reversal indicates that a main contributor to X$_{\rm CO}$ variation is the velocity dispersion typical of clouds with large clusters, which has the largest impact on the line optical depth. Finally, as shown by prior studies changing the cosmic ray ionization rate has a large impact on X$_{\rm CO}$. Increasing the cosmic ray ionization rate by a factor of 100 increases $X_{\rm CO}$, but the overall trend with efficiency and cluster mass does not change.

Finally, we show \textit{that the internal physical and chemical structure of the PDR is altered by the presence of FUV radiation from embedded forming  star clusters, with X$_{\rm CO}$ increasing by a factor of a nearly ten for smaller clusters}. High-optical depth in the CO(1-0) line reduces -- but does not eliminate -- the dependence of X$_{\rm CO}$ on the embedded (or impinging) FUV flux. We expect the change in internal physical structure has a more significant impact on optically thin tracers. The embedded flux causes an order of magnitude increase in  the internal gas temperature and significantly reduces the total molecular gas column density.  Other factors not considered in this work, including cloud sub-structure and a more distributed stellar population, will likely have a large impact -- both on $X_{\rm CO}$ and the cloud temperature distribution. We will explore these factors in future work using hydrodynamic simulations. 

\acknowledgements
The authors acknowledge helpful comments from Ron Snell, Mark Heyer, Chris McKee, Thomas Bisbas and an anonymous referee. SO acknowledges support from NSF AAG grant AST-1510021.

\bibliography{paper_bib.bib}

\begin{thebibliography}{}
\expandafter\ifx\csname natexlab\endcsname\relax\def\natexlab#1{#1}\fi

\bibitem[{{Audard} {et~al.}(2014){Audard}, {{\'A}brah{\'a}m}, {Dunham},
  {Green}, {Grosso}, {Hamaguchi}, {Kastner}, {K{\'o}sp{\'a}l}, {Lodato},
  {Romanova}, {Skinner}, {Vorobyov}, \& {Zhu}}]{audard14}
{Audard}, M., {{\'A}brah{\'a}m}, P., {Dunham}, M.~M., {et~al.} 2014, Protostars
  and Planets VI, 387

\bibitem[{{Bate} {et~al.}(2014){Bate}, {Tricco}, \&
  {Price}}]{2014MNRAS.437...77B}
{Bate}, M.~R., {Tricco}, T.~S., \& {Price}, D.~J. 2014, \mnras, 437, 77

\bibitem[{{Bell} {et~al.}(2006){Bell}, {Roueff}, {Viti}, \&
  {Williams}}]{2006MNRAS.371.1865B}
{Bell}, T.~A., {Roueff}, E., {Viti}, S., \& {Williams}, D.~A. 2006, \mnras,
  371, 1865

\bibitem[{{Bell} {et~al.}(2005){Bell}, {Viti}, {Williams}, {Crawford}, \&
  {Price}}]{2005MNRAS.357..961B}
{Bell}, T.~A., {Viti}, S., {Williams}, D.~A., {Crawford}, I.~A., \& {Price},
  R.~J. 2005, \mnras, 357, 961

\bibitem[{{Bisbas} {et~al.}(2012){Bisbas}, {Bell}, {Viti}, {Yates}, \&
  {Barlow}}]{2012MNRAS.427.2100B}
{Bisbas}, T.~G., {Bell}, T.~A., {Viti}, S., {Yates}, J., \& {Barlow}, M.~J.
  2012, \mnras, 427, 2100

\bibitem[{{Bisbas} {et~al.}(2015){Bisbas}, {Papadopoulos}, \&
  {Viti}}]{2015ApJ...803...37B}
{Bisbas}, T.~G., {Papadopoulos}, P.~P., \& {Viti}, S. 2015, \apj, 803, 37

\bibitem[{{Bisbas} {et~al.}(2017){Bisbas}, {van Dishoeck}, {Papadopoulos},
  {Sz{\H u}cs}, {Bialy}, \& {Zhang}}]{2017ApJ...839...90B}
{Bisbas}, T.~G., {van Dishoeck}, E.~F., {Papadopoulos}, P.~P., {et~al.} 2017,
  \apj, 839, 90

\bibitem[{{Bolatto} {et~al.}(2013){Bolatto}, {Wolfire}, \&
  {Leroy}}]{2013ARA&A..51..207B}
{Bolatto}, A.~D., {Wolfire}, M., \& {Leroy}, A.~K. 2013, \araa, 51, 207

\bibitem[{{Chabrier}(2005)}]{2005ASSL..327...41C}
{Chabrier}, G. 2005, in Astrophysics and Space Science Library, Vol. 327, The
  Initial Mass Function 50 Years Later, ed. E.~{Corbelli}, F.~{Palla}, \&
  H.~{Zinnecker}, 41

\bibitem[{{Chen} {et~al.}(2017){Chen}, {Braine}, {Gao}, {Koda}, \&
  {Gu}}]{2017ApJ...836..101C}
{Chen}, H., {Braine}, J., {Gao}, Y., {Koda}, J., \& {Gu}, Q. 2017, \apj, 836,
  101

\bibitem[{{Clark} \& {Glover}(2015)}]{2015MNRAS.452.2057C}
{Clark}, P.~C., \& {Glover}, S.~C.~O. 2015, \mnras, 452, 2057

\bibitem[{{Corbelli} {et~al.}(2017){Corbelli}, {Braine}, {Bandiera},
  {Brouillet}, {Combes}, {Druard}, {Gratier}, {Mata}, {Schuster}, {Xilouris},
  \& {Palla}}]{2017A&A...601A.146C}
{Corbelli}, E., {Braine}, J., {Bandiera}, R., {et~al.} 2017, \aap, 601, A146

\bibitem[{{Cubick} {et~al.}(2008){Cubick}, {Stutzki}, {Ossenkopf}, {Kramer}, \&
  {R{\"o}llig}}]{2008A&A...488..623C}
{Cubick}, M., {Stutzki}, J., {Ossenkopf}, V., {Kramer}, C., \& {R{\"o}llig}, M.
  2008, \aap, 488, 623

\bibitem[{{Draine}(1978)}]{draine78}
{Draine}, B.~T. 1978, \apjs, 36, 595

\bibitem[{{Draine}(2011)}]{2011piim.book.....D}
---. 2011, {Physics of the Interstellar and Intergalactic Medium}

\bibitem[{{Dunham} {et~al.}(2013){Dunham}, {Arce}, {Allen}, {Evans},
  {Broekhoven-Fiene}, {Chapman}, {Cieza}, {Gutermuth}, {Harvey}, {Hatchell},
  {Huard}, {Kirk}, {Matthews}, {Mer{\'{\i}}n}, {Miller}, {Peterson}, \&
  {Spezzi}}]{2013AJ....145...94D}
{Dunham}, M.~M., {Arce}, H.~G., {Allen}, L.~E., {et~al.} 2013, \aj, 145, 94

\bibitem[{{Dunham} {et~al.}(2014){Dunham}, {Stutz}, {Allen}, {Evans},
  {Fischer}, {Megeath}, {Myers}, {Offner}, {Poteet}, {Tobin}, \&
  {Vorobyov}}]{2014prpl.conf..195D}
{Dunham}, M.~M., {Stutz}, A.~M., {Allen}, L.~E., {et~al.} 2014, Protostars and
  Planets VI, 195

\bibitem[{{Enoch} {et~al.}(2008){Enoch}, {Evans}, {Sargent}, {Glenn},
  {Rosolowsky}, \& {Myers}}]{2008ApJ...684.1240E}
{Enoch}, M.~L., {Evans}, II, N.~J., {Sargent}, A.~I., {et~al.} 2008, \apj, 684,
  1240

\bibitem[{{Fischer} {et~al.}(2017){Fischer}, {Megeath}, {Furlan}, {Ali},
  {Stutz}, {Tobin}, {Osorio}, {Stanke}, {Manoj}, {Poteet}, {Booker},
  {Hartmann}, {Wilson}, {Myers}, \& {Watson}}]{2017ApJ...840...69F}
{Fischer}, W.~J., {Megeath}, S.~T., {Furlan}, E., {et~al.} 2017, \apj, 840, 69

\bibitem[{{Fletcher} \& {Stahler}(1994)}]{1994ApJ...435..313F}
{Fletcher}, A.~B., \& {Stahler}, S.~W. 1994, \apj, 435, 313

\bibitem[{{Gao} {et~al.}(2007){Gao}, {Carilli}, {Solomon}, \& {Vanden
  Bout}}]{2007ApJ...660L..93G}
{Gao}, Y., {Carilli}, C.~L., {Solomon}, P.~M., \& {Vanden Bout}, P.~A. 2007,
  \apjl, 660, L93

\bibitem[{{Glover} \& {Mac Low}(2011)}]{2011MNRAS.412..337G}
{Glover}, S.~C.~O., \& {Mac Low}, M.-M. 2011, \mnras, 412, 337

\bibitem[{{Goodman} {et~al.}(1998){Goodman}, {Barranco}, {Wilner}, \&
  {Heyer}}]{1998ApJ...504..223G}
{Goodman}, A.~A., {Barranco}, J.~A., {Wilner}, D.~J., \& {Heyer}, M.~H. 1998,
  \apj, 504, 223

\bibitem[{{Habing}(1968)}]{1968BAN....19..421H}
{Habing}, H.~J. 1968, \bain, 19, 421

\bibitem[{{Heyer} \& {Dame}(2015)}]{2015ARA&A..53..583H}
{Heyer}, M., \& {Dame}, T.~M. 2015, \araa, 53, 583

\bibitem[{{Hofner} {et~al.}(1996){Hofner}, {Kurtz}, {Churchwell}, {Walmsley},
  \& {Cesaroni}}]{1996ApJ...460..359H}
{Hofner}, P., {Kurtz}, S., {Churchwell}, E., {Walmsley}, C.~M., \& {Cesaroni},
  R. 1996, \apj, 460, 359

\bibitem[{{Hosokawa} {et~al.}(2011){Hosokawa}, {Offner}, \&
  {Krumholz}}]{hosokawa11}
{Hosokawa}, T., {Offner}, S.~S.~R., \& {Krumholz}, M.~R. 2011, \apj, 738, 140

\bibitem[{{Indriolo} {et~al.}(2007){Indriolo}, {Geballe}, {Oka}, \&
  {McCall}}]{2007ApJ...671.1736I}
{Indriolo}, N., {Geballe}, T.~R., {Oka}, T., \& {McCall}, B.~J. 2007, \apj,
  671, 1736

\bibitem[{{Indriolo} {et~al.}(2015){Indriolo}, {Neufeld}, {Gerin}, {Schilke},
  {Benz}, {Winkel}, {Menten}, {Chambers}, {Black}, {Bruderer}, {Falgarone},
  {Godard}, {Goicoechea}, {Gupta}, {Lis}, {Ossenkopf}, {Persson},
  {Sonnentrucker}, {van der Tak}, {van Dishoeck}, {Wolfire}, \&
  {Wyrowski}}]{2015ApJ...800...40I}
{Indriolo}, N., {Neufeld}, D.~A., {Gerin}, M., {et~al.} 2015, \apj, 800, 40

\bibitem[{{Kamenetzky} {et~al.}(2017){Kamenetzky}, {Rangwala}, \&
  {Glenn}}]{2017MNRAS.471.2917K}
{Kamenetzky}, J., {Rangwala}, N., \& {Glenn}, J. 2017, \mnras, 471, 2917

\bibitem[{{Kennicutt} \& {Evans}(2012)}]{2012ARA&A..50..531K}
{Kennicutt}, R.~C., \& {Evans}, N.~J. 2012, \araa, 50, 531

\bibitem[{{Krumholz} {et~al.}(2011){Krumholz}, {Klein}, \&
  {McKee}}]{2011ApJ...740...74K}
{Krumholz}, M.~R., {Klein}, R.~I., \& {McKee}, C.~F. 2011, \apj, 740, 74

\bibitem[{{Krumholz} {et~al.}(2014){Krumholz}, {Bate}, {Arce}, {Dale},
  {Gutermuth}, {Klein}, {Li}, {Nakamura}, \& {Zhang}}]{2014prpl.conf..243K}
{Krumholz}, M.~R., {Bate}, M.~R., {Arce}, H.~G., {et~al.} 2014, Protostars and
  Planets VI, 243

\bibitem[{{Kryukova} {et~al.}(2012){Kryukova}, {Megeath}, {Gutermuth},
  {Pipher}, {Allen}, {Allen}, {Myers}, \& {Muzerolle}}]{2012AJ....144...31K}
{Kryukova}, E., {Megeath}, S.~T., {Gutermuth}, R.~A., {et~al.} 2012, \aj, 144,
  31

\bibitem[{{Kryukova} {et~al.}(2014){Kryukova}, {Megeath}, {Hora}, {Gutermuth},
  {Bontemps}, {Kraemer}, {Hennemann}, {Schneider}, {Smith}, \&
  {Motte}}]{2014AJ....148...11K}
{Kryukova}, E., {Megeath}, S.~T., {Hora}, J.~L., {et~al.} 2014, \aj, 148, 11

\bibitem[{{Lagos} {et~al.}(2012){Lagos}, {Bayet}, {Baugh}, {Lacey}, {Bell},
  {Fanidakis}, \& {Geach}}]{2012MNRAS.426.2142L}
{Lagos}, C.~d.~P., {Bayet}, E., {Baugh}, C.~M., {et~al.} 2012, \mnras, 426,
  2142

\bibitem[{{Le Petit} {et~al.}(2016){Le Petit}, {Ruaud}, {Bron}, {Godard},
  {Roueff}, {Languignon}, \& {Le Bourlot}}]{2016A&A...585A.105L}
{Le Petit}, F., {Ruaud}, M., {Bron}, E., {et~al.} 2016, \aap, 585, A105

\bibitem[{{Lebr{\'o}n} {et~al.}(2011){Lebr{\'o}n}, {Mangum}, {Mauersberger},
  {Henkel}, {Peck}, {Menten}, {Tarchi}, \& {Wei{\ss}}}]{2011A&A...534A..56L}
{Lebr{\'o}n}, M., {Mangum}, J.~G., {Mauersberger}, R., {et~al.} 2011, \aap,
  534, A56

\bibitem[{{Madau} \& {Dickinson}(2014)}]{2014ARA&A..52..415M}
{Madau}, P., \& {Dickinson}, M. 2014, \araa, 52, 415

\bibitem[{{Maureira} {et~al.}(2017){Maureira}, {Arce}, {Offner}, {Dunham},
  {Pineda}, {Fernandez-Lopez}, {Chen}, \& {Mardones}}]{2017arXiv171002506M}
{Maureira}, M.~J., {Arce}, H.~G., {Offner}, S.~S.~R., {et~al.} 2017, ArXiv
  e-prints, arXiv:1710.02506

\bibitem[{{McElroy} {et~al.}(2013){McElroy}, {Walsh}, {Markwick}, {Cordiner},
  {Smith}, \& {Millar}}]{mcelroy13}
{McElroy}, D., {Walsh}, C., {Markwick}, A.~J., {et~al.} 2013, \aap, 550, A36

\bibitem[{{McKee} \& {Offner}(2010)}]{2010ApJ...716..167M}
{McKee}, C.~F., \& {Offner}, S.~S.~R. 2010, \apj, 716, 167

\bibitem[{{McKee} \& {Ostriker}(2007)}]{2007ARA&A..45..565M}
{McKee}, C.~F., \& {Ostriker}, E.~C. 2007, \araa, 45, 565

\bibitem[{{McKee} \& {Tan}(2003)}]{2003ApJ...585..850M}
{McKee}, C.~F., \& {Tan}, J.~C. 2003, \apj, 585, 850

\bibitem[{{Narayanan} \& {Hopkins}(2013)}]{2013MNRAS.433.1223N}
{Narayanan}, D., \& {Hopkins}, P.~F. 2013, \mnras, 433, 1223

\bibitem[{{Narayanan} {et~al.}(2012){Narayanan}, {Krumholz}, {Ostriker}, \&
  {Hernquist}}]{2012MNRAS.421.3127N}
{Narayanan}, D., {Krumholz}, M.~R., {Ostriker}, E.~C., \& {Hernquist}, L. 2012,
  \mnras, 421, 3127

\bibitem[{{Nelson} \& {Langer}(1997)}]{1997ApJ...482..796N}
{Nelson}, R.~P., \& {Langer}, W.~D. 1997, \apj, 482, 796

\bibitem[{{Offner} \& {Arce}(2014)}]{Offner14}
{Offner}, S.~S.~R., \& {Arce}, H.~G. 2014, \apj, 784, 61

\bibitem[{{Offner} \& {Chaban}(2017)}]{2017ApJ...847..104O}
{Offner}, S.~S.~R., \& {Chaban}, J. 2017, \apj, 847, 104

\bibitem[{{Offner} {et~al.}(2009){Offner}, {Klein}, {McKee}, \&
  {Krumholz}}]{2009ApJ...703..131O}
{Offner}, S.~S.~R., {Klein}, R.~I., {McKee}, C.~F., \& {Krumholz}, M.~R. 2009,
  \apj, 703, 131

\bibitem[{{Offner} \& {McKee}(2011)}]{2011ApJ...736...53O}
{Offner}, S.~S.~R., \& {McKee}, C.~F. 2011, \apj, 736, 53

\bibitem[{{Padovani} {et~al.}(2009){Padovani}, {Galli}, \&
  {Glassgold}}]{2009A&A...501..619P}
{Padovani}, M., {Galli}, D., \& {Glassgold}, A.~E. 2009, \aap, 501, 619

\bibitem[{{Papadopoulos}(2010)}]{2010ApJ...720..226P}
{Papadopoulos}, P.~P. 2010, \apj, 720, 226

\bibitem[{{Pineda} {et~al.}(2008){Pineda}, {Caselli}, \& {Goodman}}]{pineda08}
{Pineda}, J.~E., {Caselli}, P., \& {Goodman}, A.~A. 2008, \apj, 679, 481

\bibitem[{{R{\"o}llig} {et~al.}(2007){R{\"o}llig}, {Abel}, {Bell}, {Bensch},
  {Black}, {Ferland}, {Jonkheid}, {Kamp}, {Kaufman}, {Le Bourlot}, {Le Petit},
  {Meijerink}, {Morata}, {Ossenkopf}, {Roueff}, {Shaw}, {Spaans}, {Sternberg},
  {Stutzki}, {Thi}, {van Dishoeck}, {van Hoof}, {Viti}, \&
  {Wolfire}}]{2007A&A...467..187R}
{R{\"o}llig}, M., {Abel}, N.~P., {Bell}, T., {et~al.} 2007, \aap, 467, 187

\bibitem[{{Roman-Duval} {et~al.}(2010){Roman-Duval}, {Jackson}, {Heyer},
  {Rathborne}, \& {Simon}}]{2010ApJ...723..492R}
{Roman-Duval}, J., {Jackson}, J.~M., {Heyer}, M., {Rathborne}, J., \& {Simon},
  R. 2010, \apj, 723, 492

\bibitem[{{Rosolowsky} {et~al.}(2011){Rosolowsky}, {Pineda}, \&
  {Gao}}]{2011MNRAS.415.1977R}
{Rosolowsky}, E., {Pineda}, J.~E., \& {Gao}, Y. 2011, \mnras, 415, 1977

\bibitem[{{Rosolowsky} {et~al.}(2008){Rosolowsky}, {Pineda}, {Foster},
  {Borkin}, {Kauffmann}, {Caselli}, {Myers}, \&
  {Goodman}}]{2008ApJS..175..509R}
{Rosolowsky}, E.~W., {Pineda}, J.~E., {Foster}, J.~B., {et~al.} 2008, \apjs,
  175, 509

\bibitem[{{Safranek-Shrader} {et~al.}(2017){Safranek-Shrader}, {Krumholz},
  {Kim}, {Ostriker}, {Klein}, {Li}, {McKee}, \& {Stone}}]{2017MNRAS.465..885S}
{Safranek-Shrader}, C., {Krumholz}, M.~R., {Kim}, C.-G., {et~al.} 2017, \mnras,
  465, 885

\bibitem[{{Sandstrom} {et~al.}(2013){Sandstrom}, {Leroy}, {Walter}, {Bolatto},
  {Croxall}, {Draine}, {Wilson}, {Wolfire}, {Calzetti}, {Kennicutt}, {Aniano},
  {Donovan Meyer}, {Usero}, {Bigiel}, {Brinks}, {de Blok}, {Crocker}, {Dale},
  {Engelbracht}, {Galametz}, {Groves}, {Hunt}, {Koda}, {Kreckel}, {Linz},
  {Meidt}, {Pellegrini}, {Rix}, {Roussel}, {Schinnerer}, {Schruba}, {Schuster},
  {Skibba}, {van der Laan}, {Appleton}, {Armus}, {Brandl}, {Gordon}, {Hinz},
  {Krause}, {Montiel}, {Sauvage}, {Schmiedeke}, {Smith}, \&
  {Vigroux}}]{2013ApJ...777....5S}
{Sandstrom}, K.~M., {Leroy}, A.~K., {Walter}, F., {et~al.} 2013, \apj, 777, 5

\bibitem[{{Seifried} \& {Walch}(2016)}]{2016MNRAS.tmpL..19S}
{Seifried}, D., \& {Walch}, S. 2016, \mnras, arXiv:1510.06544

\bibitem[{{Sembach} {et~al.}(2000){Sembach}, {Howk}, {Ryans}, \&
  {Keenan}}]{2000ApJ...528..310S}
{Sembach}, K.~R., {Howk}, J.~C., {Ryans}, R.~S.~I., \& {Keenan}, F.~P. 2000,
  \apj, 528, 310

\bibitem[{{Shetty} {et~al.}(2011){Shetty}, {Glover}, {Dullemond}, \&
  {Klessen}}]{2011MNRAS.412.1686S}
{Shetty}, R., {Glover}, S.~C., {Dullemond}, C.~P., \& {Klessen}, R.~S. 2011,
  \mnras, 412, 1686

\bibitem[{{Shu}(1977)}]{1977ApJ...214..488S}
{Shu}, F.~H. 1977, \apj, 214, 488

\bibitem[{{Shu} {et~al.}(1995){Shu}, {Najita}, {Ostriker}, \& {Shang}}]{shu95}
{Shu}, F.~H., {Najita}, J., {Ostriker}, E.~C., \& {Shang}, H. 1995, \apjl, 455,
  L155

\bibitem[{{Soderblom} {et~al.}(2014){Soderblom}, {Hillenbrand}, {Jeffries},
  {Mamajek}, \& {Naylor}}]{2014prpl.conf..219S}
{Soderblom}, D.~R., {Hillenbrand}, L.~A., {Jeffries}, R.~D., {Mamajek}, E.~E.,
  \& {Naylor}, T. 2014, Protostars and Planets VI, 219

\bibitem[{{Sodroski} {et~al.}(1995){Sodroski}, {Odegard}, {Dwek}, {Hauser},
  {Franz}, {Freedman}, {Kelsall}, {Wall}, {Berriman}, {Odenwald}, {Bennett},
  {Reach}, \& {Weiland}}]{1995ApJ...452..262S}
{Sodroski}, T.~J., {Odegard}, N., {Dwek}, E., {et~al.} 1995, \apj, 452, 262

\bibitem[{{Spaans} \& {van Dishoeck}(1997)}]{1997A&A...323..953S}
{Spaans}, M., \& {van Dishoeck}, E.~F. 1997, \aap, 323, 953

\bibitem[{{Strong} {et~al.}(2004){Strong}, {Moskalenko}, {Reimer}, {Digel}, \&
  {Diehl}}]{2004A&A...422L..47S}
{Strong}, A.~W., {Moskalenko}, I.~V., {Reimer}, O., {Digel}, S., \& {Diehl}, R.
  2004, \aap, 422, L47

\bibitem[{{Tan} {et~al.}(2014){Tan}, {Beltr{\'a}n}, {Caselli}, {Fontani},
  {Fuente}, {Krumholz}, {McKee}, \& {Stolte}}]{2014prpl.conf..149T}
{Tan}, J.~C., {Beltr{\'a}n}, M.~T., {Caselli}, P., {et~al.} 2014, Protostars
  and Planets VI, 149

\bibitem[{{Tout} {et~al.}(1996){Tout}, {Pols}, {Eggleton}, \&
  {Han}}]{1996MNRAS.281..257T}
{Tout}, C.~A., {Pols}, O.~R., {Eggleton}, P.~P., \& {Han}, Z. 1996, \mnras,
  281, 257

\bibitem[{{Vasyunina} {et~al.}(2014){Vasyunina}, {Vasyunin}, {Herbst}, {Linz},
  {Voronkov}, {Britton}, {Zinchenko}, \& {Schuller}}]{2014ApJ...780...85V}
{Vasyunina}, T., {Vasyunin}, A.~I., {Herbst}, E., {et~al.} 2014, \apj, 780, 85

\bibitem[{{Vaytet} \& {Haugbolle}(2016)}]{vaytet16}
{Vaytet}, N., \& {Haugbolle}, T. 2016, VizieR Online Data Catalog, 359

\bibitem[{{Walch} {et~al.}(2015){Walch}, {Girichidis}, {Naab}, {Gatto},
  {Glover}, {W{\"u}nsch}, {Klessen}, {Clark}, {Peters}, {Derigs}, \&
  {Baczynski}}]{2015MNRAS.454..238W}
{Walch}, S., {Girichidis}, P., {Naab}, T., {et~al.} 2015, \mnras, 454, 238

\bibitem[{{Wolfire} {et~al.}(2010){Wolfire}, {Hollenbach}, \&
  {McKee}}]{2010ApJ...716.1191W}
{Wolfire}, M.~G., {Hollenbach}, D., \& {McKee}, C.~F. 2010, \apj, 716, 1191

\bibitem[{{Yun} {et~al.}(2015){Yun}, {Aretxaga}, {Gurwell}, {Hughes},
  {Monta{\~n}a}, {Narayanan}, {Rosa-Gonz{\'a}lez}, {S{\'a}nchez-Arg{\"u}elles},
  {Schloerb}, {Snell}, {Vega}, {Wilson}, {Zeballos}, {Chavez}, {Cybulski},
  {D{\'{\i}}az-Santos}, {De La Luz}, {Erickson}, {Ferrusca}, {Gim}, {Heyer},
  {Iono}, {Pope}, {Rogstad}, {Scott}, {Souccar}, {Terlevich}, {Terlevich},
  {Wilner}, \& {Zavala}}]{2015MNRAS.454.3485Y}
{Yun}, M.~S., {Aretxaga}, I., {Gurwell}, M.~A., {et~al.} 2015, \mnras, 454,
  3485

\end{thebibliography}

\appendix
\section{Model Variations}\label{appendix:facc}

In this appendix, we revisit Figure \ref{fig:meffxfac} and discuss the impact of our accretion model assumptions. Our fiducial model, TTC, agrees well with observations of larger clusters, but it over-predicts the luminosities of some of the smaller clusters. This disagreement mainly applies to cluster data from \citet{2012AJ....144...31K}, since these include a larger number of low-luminosity sources. 

There are several possible explanations for this discrepancy. First, our luminosity formalism could be inaccurate. The model includes several tunable factors, including the accretion coefficient, $\dot m_0$ and the fraction of accretion energy radiated away, $f_{\rm acc}$. The former parameter depends on local physical parameters such as the column density or temperature, which vary from region to region. The latter parameter is uncertain since it depends on pre-main sequence model assumptions and the outflow/wind launching mechanism  \citep{hosokawa11,shu95}. However, $f_{\rm acc}$ is estimated to be between 0.5 and 1, which is a relatively narrow range of uncertainty. A more significant uncertainty underpines the choice of protostellar radii. These are debated to factors of two, although some authors have argued that the initial radii are largely independent of stellar mass \citep{hosokawa11,vaytet16} and the evolution is insensitive to the accretion history \citep{hosokawa11}.

A more comprehensive concern is the form of the accretion model, which may be incorrect. The TC model was formulated for massive stars ($M \gtrsim 10$ $M_\odot$) and may simply not represent smaller clusters, which are dominated by lower mass stars.  To address this, \citet{2011ApJ...736...53O} proposed the two-component turbulent core model (2CTC), which allows for lower mass cores in which turbulent pressure is comparable to or smaller than the thermal pressure. However, this hybrid formalism shifts the peak of the PMF and PLF to slightly higher masses and luminosities, respectively; adopting tapered 2CTC in lieu of TTC would increase disagreement between the models and observations of small clusters. Alternatively, the competitive accretion (CA) model,  as adapted by  \citet{2011ApJ...736...53O}, predicts lower typical luminosities. In fact, \citet{2012AJ....144...31K} found that the CA model exhibited the best agreement with their data. However, this model would produce an overall shift to lower luminosities, potentially reducing the agreement between the models and higher mass clusters.

\begin{figure*}
\centering
\includegraphics[width=\textwidth]{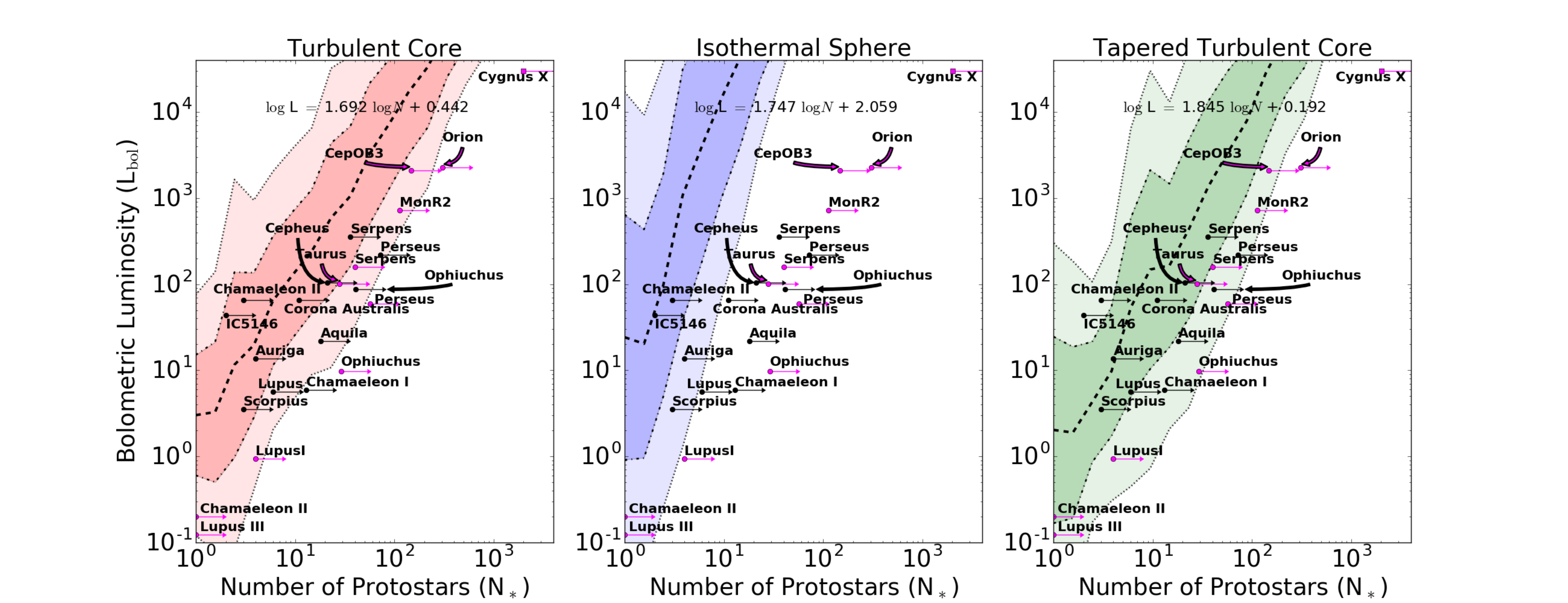}
\caption{\label{fig:clusterfacc05}Total cluster luminosity versus the number of protostars in the galaxy for three different accretion histories. The black solid lines indicate the mean of the luminosity distributions. The dark and light colored bands indicate the 1 and 2 $\sigma$ spread of the distribution. The black data points indicate the sum of the bolometric luminosities for each cluster in \cite{2013AJ....145...94D}. The pink circles show clusters from the \cite{2012AJ....144...31K} catalog and the pink square is Cygnus X from the \cite{2014AJ....148...11K}. The best fit to the mean total luminosity is annotated on each plot.}
\end{figure*}

A final possibility is that accretion may be variable or episodic \citep[][and references therein]{audard14}. One way to account for episodic accretion is by modifying $f_{\rm acc}$. If $f_{\rm epi}$ is the fraction of mass accreted during episodic events, then the effective $f_{\rm acc}$ can be written \citep{2011ApJ...736...53O}:
\begin{equation}
f_{\rm acc, eff} = f_{\rm acc} \left ( 1 - f_{\rm epi} \right )
\end{equation}
Note that this formulation implicitly assumes that accretion bursts are rare and short-lived. In this case, episodic events are likely absent in small statistical samples, such as those representative of Gould Belt clouds. Through comparisons with mean protostellar luminosities in local regions \citet{2011ApJ...736...53O} suggested an effective value of $f_{\rm acc} = 0.56$. Figure \ref{fig:clusterfacc05} shows the bolometric luminosity predictions for our three accretion models with $f_{\rm acc, eff} = 0.5$. This value corresponds to $f_{\rm epi} = \frac{1}{3}$. The total luminosities are lower, and more moderately sized clusters fall within the 2$\sigma$ bounds. In fact, some degree of episodic accretion could explain why discrepancies appear with smaller clusters but not more massive ones: more massive clusters are sufficiently well-sampled to include some bursts.
 
In conclusion a great deal of uncertainty underlies protostellar accretion. Different models may produce degenerate results as noted by \citet{2014prpl.conf..195D}, and additional constraints are needed to converge on the most accurate model.

\section{Tapering Parameter} \label{appendix:tappar}
\begin{figure*}
\plotone{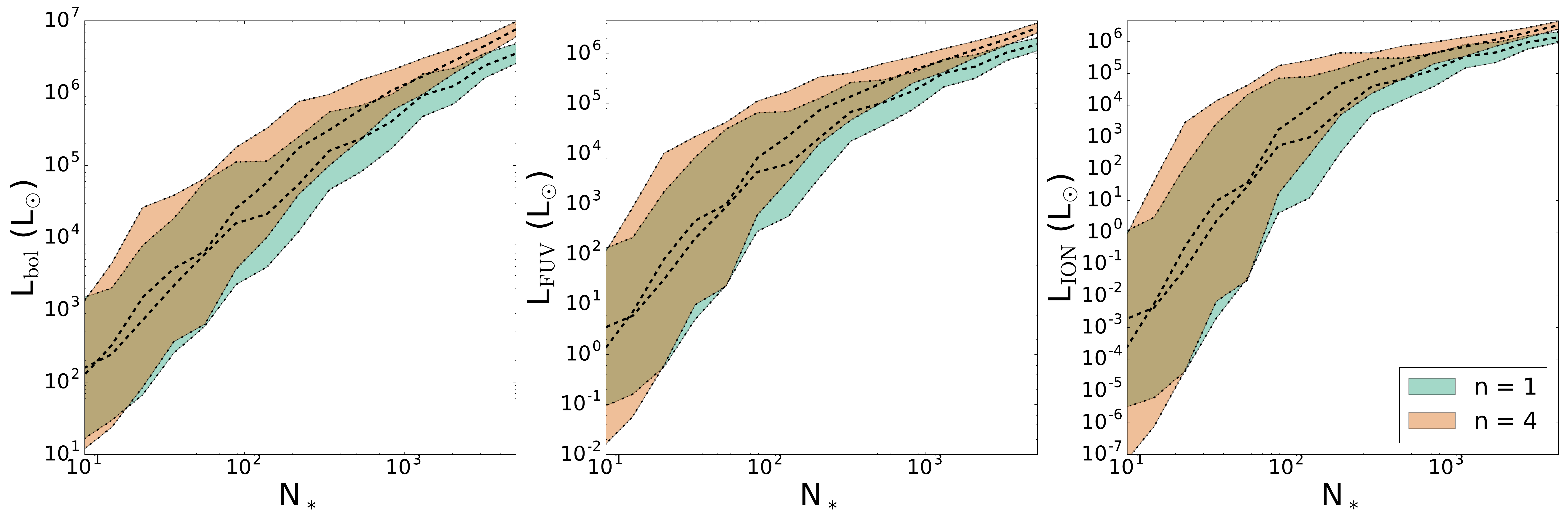}
\caption{\label{fig:tapercomp}Cluster luminosity as a function of the number of protostars in the cluster for the $n = 1$ and $n = 4$ tapering models. The left panel is the bolometric luminosity, the center is the FUV luminosity and the right is the ionizing luminosity.}
\end{figure*}

\cite{2010ApJ...716..167M} adopts tapered accretion histories with a general form of
\begin{equation}
\dot{m} = \dot{m}_1 \left ( \frac{m}{\mf} \right )^j \mf^{j_f} \left [ 1 - \left ( \frac{t}{t_f} \right )^n \right ],
\end{equation}
where $n$ defines how steeply the accretion tapers. In this work, we use $n = 1$, such that the formation time of stars is twice that of stars with untapered accretion. Recent magneto-hydrodyamic simulations of isolated star-forming cores by \cite{2017ApJ...847..104O} indicate $n = 4$. Figure \ref{fig:tapercomp} shows the cluster luminosities as a function of the number of protostars for both the $n = 1$  and $n = 4$ tapering cases. For smaller clusters the luminosities are similar within the spread. For larger clusters the $n = 4$ clusters are brighter by a factor of few. The exact form of the accretion history is poorly constrained by observations \citep{2014prpl.conf..195D}, although there is some support for steeper tapering for protostars in Orion \citep{2017ApJ...840...69F}. Given the spread of bolometric luminosities, a much larger statistical sample of clusters would be needed to better constrain the tapering parameter, $n$. 

\section{Time Dependence and Main Sequence Stars}\label{appendix:msstars}
Star formation within a cloud is not instantaneous and is likely spread over a few million years \citep{2014prpl.conf..219S}. Thus, for a given cloud the first forming stars will be on the main sequence (MS) by the time the last generation of protostars appears. Here, we assume that star formation occurs at a steady state and that all the stellar objects contributing to the total luminosity are still protostars. However, this is an approximation, which is most accurate for young clusters less than $\sim 1$ Myr old. In this appendix we investigate the impact of an additional population of MS stars on the total bolometric luminosity.

\cite{1994ApJ...435..313F} modeled evolving star clusters assuming an IS accretion model and followed the populations of both protostars and MS stars.  This naturally produces a time dependent cluster luminosity. For our work, this suggests an additional degree of freedom for $X_{\rm CO}$: the cluster age. Since stars with different star masses have different formation times, this implies that not only the number but the mass distribution of MS stars is a strong function of age and the accretion model. At early times, however, most of the cluster members are still protostars. To assess the impact of MS stars on the cluster luminosities,  we generate mock clusters where instead of sampling from the bivariate PMF we draw the populations from the IMF, i.e., we assume all the stars are on the MS.  The luminosities and radii of the MS stars are from \cite{1996MNRAS.281..257T}. Such clusters represent an idealized case where star formation has recently ended. For comparison, we also generate mock clusters with twice as many stars but where half of the population are protostars sampled from the bivariate PMF and the other half are MS stars. This approximates clusters at an intermediate time of their formation.

\begin{figure}
\centering
\plotone{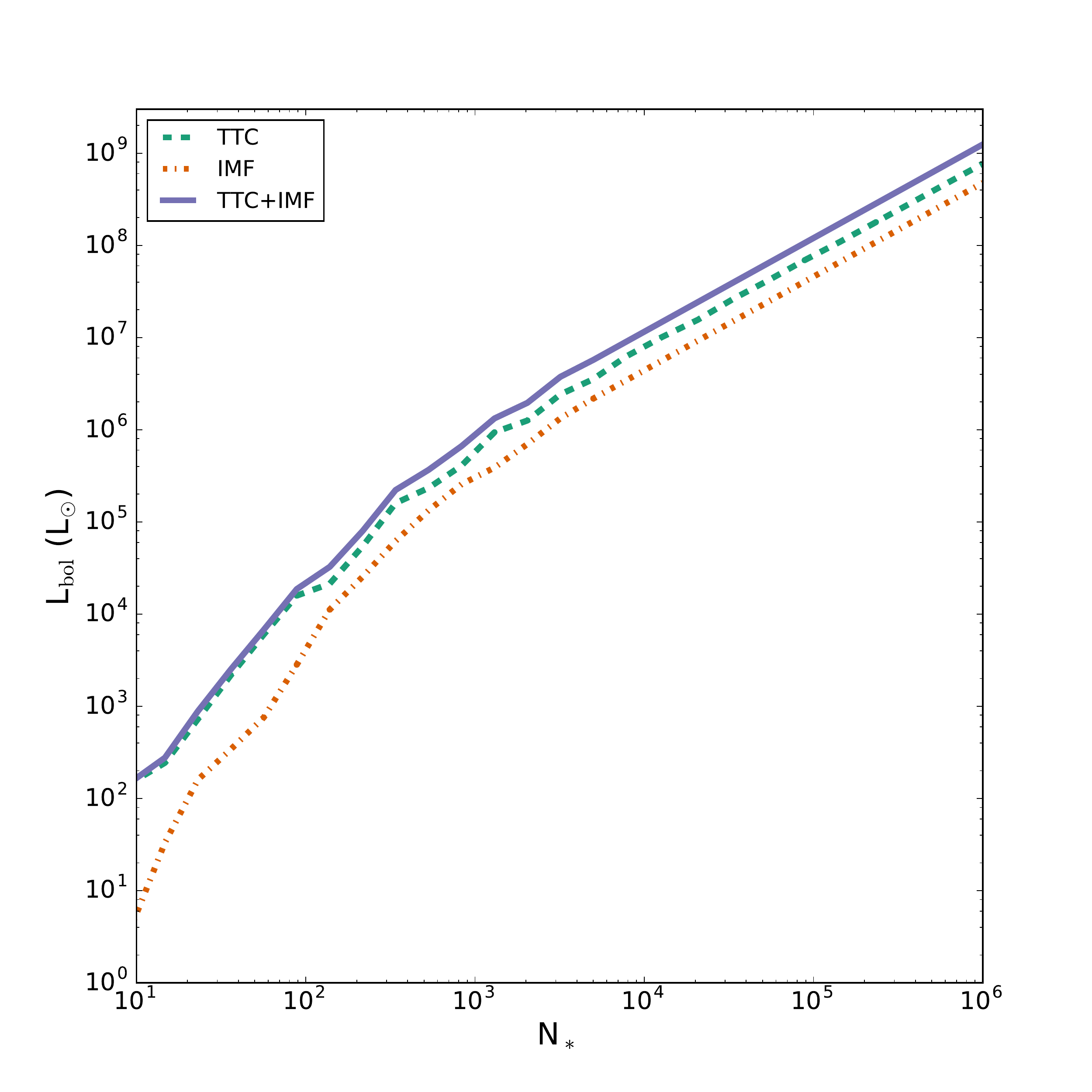}
\caption{\label{fig:pmfvsimf} Total cluster luminosity as a function of the number of members in the cluster. The blue dashed line is the mean luminosity for the TTC model, the orange dashed-dot line for the IMF model and the solid purple line for the IMF+TTC combined model.}
\end{figure}

Figure \ref{fig:pmfvsimf} shows the mean cluster bolometric luminosity as a function of the number of cluster members. Clusters composed entirely of MS stars have lower luminosities compared to their protostar counterparts. This is especially true for small clusters, where accretion luminosity dominates. Larger clusters composed of N$_*$ protostars and N$_*$ MS stars have luminosities that are higher by a factor of two. This difference is driven by the dominance of higher mass stars, whose internal luminosity exceeds their accretion luminosity. Therefore, assuming the clusters we model are relatively young, we expect a secondary MS population to have minimal impact on our conclusions for small clusters. In contrast, our models may underestimate the true luminosities of large clusters that are somewhat advanced in star formation by a factor of $\sim$2. 

\end{document}